\documentclass[11pt,fleqn,twoside,titlepage]{newcslarticle}
\cslreportnumber{CSL Technical Report SRI-CSL-25-01R2}
\acknowledge{SRI Project 101425 in support of
DARPA ANSR Program.\\
Distribution Statement ``A'' (Approved for Public Release, Distribution Unlimited).}
\usepackage{fancyhdr}
\usepackage{graphicx,caption,cite,url,relative,latexsym,alltt,nobibhead}
\PassOptionsToPackage{hyphens}{url}
\usepackage[bookmarks=true,hyperfigures=true,colorlinks=true,linkcolor=blue,citecolor=blue,backref=page,pagebackref=false,pdfpagemode=fullscreen,plainpages=false,pdfpagelabels]{hyperref}
\usepackage[all]{hypcap}
\usepackage[normalem]{ulem} 
\usepackage{caption,cite,url,latexsym,alltt}
\usepackage[utf8]{inputenc}
\usepackage{array}
\usepackage{xspace}

\usepackage{wrapfig}

\def\BigLaTeX{{\rm L\kern-.36em\raise.3ex\hbox{\smaller\smaller A}\kern-.15em
    T\kern-.1667em\lower.7ex\hbox{E}\kern-.125emX}}
\def\BoldLaTeX{{\bf L\kern-.36em\raise.3ex\hbox{\smaller\smaller\bf A}\kern-.15em
    T\kern-.1667em\lower.7ex\hbox{E}\kern-.125emX}}
\def\BibTeX{{\rm B\kern-.05em{\sc i\kern-.025em b}\kern-.08em
    T\kern-.1667em\lower.7ex\hbox{E}\kern-.125emX}}

\newlength{\hsbw}
\def\extrawidth{0.5in}

\newlength{\vsbw}
\newcounter{sessioncount}
\newenvironment{session*}{\begin{flushleft}
 \refstepcounter{sessioncount}
 \setlength{\hsbw}{\linewidth}
 \addtolength{\hsbw}{-\arrayrulewidth}
 \addtolength{\hsbw}{-\tabcolsep}
 \begin{tabular}{@{}|c@{}|@{}}\hline 
 \begin{minipage}[b]{\hsbw}
 \vspace*{-.5pt}
 \begin{flushright}
 \rule{0.01in}{.15in}\rule{0.3in}{0.01in}\hspace{-0.35in}
 \raisebox{0.04in}{\makebox[0.3in][c]{\footnotesize \thesessioncount}}
 \end{flushright}
 \vspace*{-.57in}
 \begingroup\small\vspace*{1.0ex}\begin{alltt}}{\end{alltt}\endgroup\end{minipage}\\ \hline 
 \end{tabular}
 \end{flushleft}}
\def\sessionsize{\small}
\def\smallsessionsize{\small}

\newcommand{\exmemo}[1]{}

\newcommand{\comment}[1]{}

\newcommand{\exfootnote}[1]{}
\newlength{\sblen}
\newlength{\overhang}

\def\SetFigFont#1#2#3{\rm}

\newcommand{\excite}[1]{}

\newcommand{\arxiv}[1]{\href{https://arxiv.org/abs/#1}{\tt arXiv:#1}}
\newcommand{\doi}[1]{DOI = \url{https://doi.org/#1}}

\newtheorem{theorem}{Theorem}

\def\min{\mathop{\rm min}}

\newcommand{\aless}{\mathrel{\mbox{\lower.9ex\hbox{$\stackrel{\textstyle <}{\sim}$}}}}
\newcommand{\amore}{\mathrel{\mbox{\lower.9ex\hbox{$\stackrel{\textstyle >}{\sim}$}}}}

\newcommand{\blue}[1]{\textcolor{blue}{ #1}}
\newcommand{\vbar}{\,|\,}
\newcommand{\mult}{\(\times\)}
\newcommand{\aaa}{\(\wedge\)}
\newcommand{\imp}{\(\supset\) }
\newcommand{\pp}{$^\prime\!$}
\newcommand{\xx}{\(\times\)}

\newcommand{\ii}{\(\cap\)}

\newcommand{\ok}{\mathrm{ok}}
\newcommand{\doubt}{concern\xspace }
\newcommand{\doubts}{concerns\xspace }

\newcommand{\Doubts}{Concerns\xspace }
\newcommand{\ifelse}[2]{#2}

\title{\bf Quantifying Confidence\\[1ex] In Assurance 2.0 Arguments} 
\author{Robin Bloomfield (City St George's, University\ of London) and\\[0.5ex]John Rushby (SRI)}
\date{29 December 2025}
\begin{document}
\maketitle
\begin{abstract}
\begin{quotation}

Confidence is central to safety and assurance cases: how much
confidence a decision requires and how much the argument actually
provides are both important questions.  We present a new method for
assessing probabilistic confidence in assurance case arguments that is
simple, systematic and sound.  

It exploits the ways claims are decomposed in a structured argument
and provides different approaches according to the different degrees
of (in)dependence and diversity among subclaims and the way they
eliminate concerns that undermine confidence in their parent
claims. The method uses only elementary probabilistic constructions
that are well-known in other contexts (e.g., Fr\'{e}chet bounds) but
we interpret and apply them in a manner that is specifically focused
on assurance arguments and requires no background in probabilistic
analysis.

We show that the method is not susceptible to the counterexamples that
Graydon and Holloway exhibit for other approaches to quantifying
confidence \cite{Graydon&Holloway:quant17} and we recommend it as an
additional tool in assessment of Assurance 2.0 arguments.  The primary
evaluation criteria for Assurance 2.0 remain logical indefeasibility
and dialectical examination, but probabilistic assessment can be
useful in evaluating cost/confidence tradeoffs for different risk
levels, and the overall balance of confidence across a structured
argument.

\end{quotation}
\end{abstract}

\setlength{\mathindent}{1.6em}

\newpage
\setcounter{page}{1}
\pagenumbering{roman}

\tableofcontents
\clearpage
\listoffigures

\listoftables

\clearpage
\setcounter{page}{1}
\pagenumbering{arabic}

\renewcommand{\sectionmark}[1]{\markboth{\thesection.\ #1}{\thesection.\ #1}}
\renewcommand{\subsectionmark}[1]{\markright{\thesubsection.\ #1}}
\renewcommand{\headrulewidth}{0pt}
\setlength{\headheight}{13.6pt}

\fancyhead[l]{\iffloatpage{}{\rightmark}}
\fancyhead[c]{}
\fancyhead[r]{}

\pagestyle{fancy}
\section{Introduction}

We have previously described overall assessment of confidence in an
assurance case \cite{Bloomfield&Rushby24:CBJ} within the methodology
that we call Assurance 2.0 \cite{Bloomfield&Rushby:Assurance2}, with
more details provided in a supporting technical report
\cite{Bloomfield&Rushby:confidence22}.  The primary positive
assessment is \emph{logical soundness}, and this should be subjected
to Socratic/dialectical challenge by exploration of potential
\emph{defeaters} \cite{Bloomfield-etal:defeaters24}.  For overall
confidence, all defeaters must themselves be defeated or else accepted
as \emph{residual doubts} that have been shown to pose negligible
risk, and soundness must be established \emph{indefeasibly}
\cite{Rushby:Shonan16}, meaning no credible new information would
change the assessment.

But indefeasible assurance does not amount to certainty: there
can always be some \emph{aleatoric uncertainty} (uncertainty
\emph{in} the world, such as how many sensors fail during a mission)
and some \emph{epistemic uncertainty} (uncertainty \emph{about} the
world, such as whether our evidence for some attribute is really
conclusive).  These can be reduced (at increased cost) but not
eliminated, so we set a threshold that balances cost and
confidence.  Often, this balance is struck and stated informally (if
stated at all), but sometimes it can be useful to attempt to quantify
and express it numerically.

Some authors are dubious of quantification because they consider that
it conveys a false sense of precision \cite{Graydon&Holloway:quant17}.
A contrary opinion asserts that quantification can support explicit
reasoning and structured decision-making.  Decision makers need to
evaluate choices among the alternatives with care, cognizant of
their strengths and weaknesses.  For example, Spiegelhalter
\cite{Spiegelhalter04:art} notes that in decision-making for the Bay
of Pigs Invasion there were different interpretations for
the qualitative assessment of ``fair chance'' that contributed to a
major strategic failure.  But there are also instances of egregious
decision making based on unjustifiably precise quantified risk
assessments.  The Challenger disaster is an example: engineers had
serious qualitative concerns about O-ring resilience at low
temperatures, yet management relied on quantitative risk estimates
that underestimated epistemic uncertainty and ignored the probability
of catastrophic failure.  These examples illustrate the central
tension in quantified statements of assurance: used carelessly or
without adequate grounding, numbers can suppress judgment, but when
used appropriately they can also sharpen thinking and assist decision
making.

Quantification can be particularly valuable in situations where a
graduated treatment of confidence is needed.  This arises in systems
where different items pose different risks, and is exemplified by the
Design Assurance Levels (DALs) A--E of DO-178C for software in
commercial aircraft \cite{DO178C}, the Safety Integrity Levels (SILs)
4--1 of IEC 61508 for Programmable Electronic Systems \cite{IEC61508},
and the Automotive SILs (ASILs) D--A of ISO 26262 for cars
\cite{26262} (in each case we list the levels in descending order).
In Assurance 2.0 we insist on logical soundness, even for items that
pose less risk, but to reduce the cost of their assurance we may
reduce some thresholds on evidence and rigor.  To make the reductions
in a rational and justifiable manner, it is useful to assess some
numerical estimate of confidence in an assurance case, and that is the
topic of this report.

Subjective probability \cite{Jeffrey04} provides a natural measure for
numerical assessment of confidence: that is, confidence in a claim
\texttt{X} is our subjective estimate that is it true, expressed as a
probability, \texttt{P(X)}.  Hence, our numerical assessment of
confidence will use ideas and techniques from probability theory.
Note that is fundamental that confidence assessments for a semi-formal
assurance case are subjective judgments, preferably achieved through
consensus using rational methods of assessment and calculation: they
are not frequentist or other objective measurements but can be
combined with such measures (e.g., absence of failure in system
testing or early deployment) to yield credible worst-case estimates of
long-term reliability \cite{Strigini&Povyakalo13}.

We expect the numerical assessment to be \emph{compositional}: that
is, built up step by step, ascending the structured assurance case
argument from its leaves (typically, evidence) to its top claim (the
conclusion).  In such cases, numerical assessment can also be used to
examine the distribution of confidence across the argument, the
sensitivity of confidence in the conclusion to that of evidence and
intermediate steps, and to compare one argument with another.

As explained previously \cite{Bloomfield&Rushby:Assurance2},
we have two ways of organizing assessment of probabilistic confidence.
\begin{itemize}

\item There are circumstances where some probabilistic measure (e.g.,
probability of failure on demand, \emph{pfd}) is an explicit part of
the top claim (some nuclear systems are like this).  In this case,
probabilistic measures will also appear in the evidence assembled and
in the internal claims developed within the argument, and these
measures will be propagated by specific theories that are cited by the
steps of the argument.  We refer to this as \emph{internal}
probabilistic assessment.

\item In other cases, the top claim will be unconditional (e.g., ``the
system is correct''), and we develop a separate \emph{external}
probabilistic assessment of our subjective confidence (i.e., belief)
in the truth of this claim and of those claims and evidence that
support it.
\end{itemize}
(It is plausible that these methods could be used in combination: for
example, aleatoric uncertainty could be treated internally while
epistemic uncertainty is assessed externally.)

Here, we are concerned with external assessments, where we and many
others have proposed methods for calculating probabilistic or other
numerical assessments.  Most of these methods assume that assurance
cases are organized around a structured argument that relates evidence
about the system to significant claims about it (others use Toulmin
arguments, Bayesian Belief Networks, or other diagrammatic
representations).  The argument must pass some threshold for logical
validity, soundness, and persuasiveness but the rigor of this
assessment varies according to the method.  Most methods other than
Assurance 2.0 apply only informal assessments of logical correctness
and their techniques for numerical assessment therefore carry some, or
much, of the burden for overall assessment of veridity.  In Assurance
2.0 on the other hand, we assess logical soundness rigorously,
separately, and prior to, numerical assessment.  This allows our
numerical assessment to focus on providing a view that is supportive
and complementary to the logical one, and enables simple probabilistic
models because they serve only a single purpose.  However, notice that
because it provides a different view, numerical assessment contributes
to dialectical examination and can raise issues that cause logical
assessment to be revisited.  For example, we may initially make
deterministic aleatoric assumptions about failure modes or event
occurrences where a probabilistic treatment shows a more nuanced
approach is to be preferred.

Our previous probabilistic treatments
\cite{Bloomfield&Rushby24:CBJ,Bloomfield&Rushby:confidence22} used
very elementary and conservative calculations.  We briefly summarize
these treatments in the following section and then, in Section
\ref{method}, we introduce our new method and its application to
several different kinds of \emph{reasoning} steps within an argument.
Our new method is very simple and uses only classical probabilistic
reasoning.  Section \ref{method} is the main novel contribution in
this report.  However, in its final example, we show that our previous
treatment is the correct choice for certain reasoning steps and from
there we segue in Section \ref{bbns} to a treatment for similar but
more complex examples using Bayesian Belief Networks (BBNs).  We have
covered this topic before \cite{Rushby:AAA15}, but the presentation
here is integrated into our larger narrative.  In Section
\ref{compare} we compare our new treatment with some other methods for
quantifying confidence in assurance cases.  These use different
probabilistic models, including those derived from theories of
evidence such as Dempster-Shafer, and alternative applications of
BBNs.  In Section \ref{GandH} we consider the critique provided by
Graydon and Holloway
\cite{Graydon&Holloway:quant17,Graydon&Holloway:quant16} for some of
these other methods and explain why ours are not susceptible to them.
We provide conclusions in Section \ref{conc}.

We conclude this introduction with a brief overview of Assurance 2.0
for those who have not yet read the earlier documents.

We model an assurance case as a network of argument steps over
propositions called \emph{claims}, represented diagrammatically as
\emph{claim nodes}, connected by \emph{argument nodes} or
\emph{steps}.  A claim is intended to denote an atomic proposition
(typically expressed in controlled natural language) or a conjunction
or disjunction of these.  An argument step relates a parent claim to a
set of supporting \emph{subclaims} that function as premises, together
with an optional \emph{sideclaim} that records any explicit conditions
required for the step to be sound.  The combination of an argument
step and its parent claim, subclaims, and sideclaim is called an
\emph{argument block} (as in building block).  Assurance 2.0 has just
six different types of block
\cite{Bloomfield&Netkachova14}.\footnote{Exact defeater blocks are an
addition to the original five.}

We distinguish subclaims, which are the principal supporting premises
offered for the parent claim, from the \emph{sideclaim}, which
captures the \emph{applicability conditions} for the warrant or
\emph{justification} for the argument scheme used in the step (e.g.,
assumptions about scope, model fidelity, tool qualification,
independence, or completeness etc.).  This distinction is
methodological rather than logical---sideclaims could be represented
as ordinary premises---but it is useful because such conditions on the
warrant are often the main source of residual doubt and are best
scrutinized explicitly.\footnote{\label{resids}Historically, we have spoken of
residual \emph{doubts}, but in a probabilistic context we prefer to
speak of residual \emph{concerns} as \emph{doubt} is used for the
probabilistic function $1 - P(a)$.}

In Assurance~2.0, we treat the assessment process as layered: (i)
logical scrutiny addresses coherence and logical validity, then
evidence and warrant suitability justify soundness; (ii) dialectical
scrutiny examines potential defeaters; and only then (iii) a
quantitative confidence layer examines remaining epistemic uncertainty
by assigning assessor confidence to claims and (where relevant) to
inference strength.  Confidence is represented as a subjective
probability, and the aim is not to turn assurance into statistical
sampling, but to make residual uncertainty explicit and compositional,
enabling sensitivity analysis and principled trade-offs between
evidence-gathering cost and achieved confidence.

\section{Our Previous Method}
\label{oldmethod}

Our previous treatments
\cite{Bloomfield&Rushby24:CBJ,Bloomfield&Rushby:confidence22} used
very elementary calculations based on probabilistic logic.  The
logical interpretation of an Assurance 2.0 argument treats each block
as a subargument with its parent claim as the conclusion and its
subclaims and sideclaim (if any) as the premises.  The premises are
implicitly conjoined and the standard interpretation of conjunction in
probabilistic logics \cite{Adams98,Nilsson:prblog86} is a product:
\texttt{P(A \& B) = P(A)\,\mult\,P(B\vbar A)}, which can be
simplified, if \texttt{A} and \texttt{B} are independent, to
\texttt{P(A)\mult{}P(B)}.  We argued that because all claims and
subclaims must be true in an assurance argument, the simple form is
sound and probabilistic confidence can therefore be propagated upwards
from the leaf nodes (i.e., from confidence in evidence) by this
\emph{product calculation}, where confidence in each claim is the
product of confidence in its subclaims (and sideclaim, if any).
Hence, when \texttt{conf(X)} denotes confidence in claim \texttt{X},
the general product expression (neglecting any sideclaim) for a
decomposition block with $n$ subclaims is
\[
\texttt{conf(parent)} \geq \prod_{i=1}^n \texttt{conf(subclaim$_i$)}.
\]

We also showed that a more conservative calculation, which applies
with correlated claims, propagates confidence as the \emph{sum of
doubts}, where \texttt{doubt(X) = 1 - conf(X)} and doubt in each claim
is calculated as the sum of doubts in its subclaims.  Thus, the
general expression for a decomposition block with $n$ subclaims
(again, neglecting any sideclaim) is
\[
\texttt{doubt(parent)} \leq \min\left(1, \sum_{i=1}^n \texttt{doubt(subclaim$_i$)}\right).
\]
This can also be written as
\[
\texttt{conf(parent)} \geq \max\left(0, \sum_{i=1}^n \texttt{conf(subclaim$_i$)} - (n-1)\right)
\]
where the right hand side is known as the \emph{Fr\'{e}chet lower
bound} for intersection \cite{frechet-wiki}.

Both the product and the sum of doubts calculations produce highly
conservative values that decrease as they go higher in the argument
tree (so calculated confidence in the top claim is always very low)
and are largely indifferent to its shape (i.e., to the actual
argument).  Consequently, they are of limited utility and we no longer
recommend them for general use, although they are appropriate in some
circumstances (e.g., see Section \ref{cumulative}).

This limited utility is not surprising in retrospect: logic and
probability have different purposes and model the world differently,
so no combination of the two works well in all circumstances.  There
can be many reasons why an argument step has several subclaims: for
example, different subclaims could address different circumstances, or
they could address the same circumstance in diverse ways (so that
confidence may increase as we ascend the argument) and a probabilistic
assessment should treat each according to its reasons, whilst logical
assessment always lumps all subclaims together as a conjunction and
simply requires that each is \texttt{true}.  Hence, in the new
treatment, and unlike our previous ones, we largely separate logical
and probabilistic interpretations although the latter assumes the
former has been applied satisfactorily.

\section{The New Method}
\label{method}

Our new method for assessing probabilistic confidence in an Assurance
2.0 case works compositionally, claim by claim, on the argument of the
case from bottom to top.

The leaf nodes at the bottom of a completed assurance case argument
can be references to external subcases, assumptions, evidence, or
residual doubts.\footnote{An incomplete argument may also contain
defeaters, or explicitly unfinished nodes.  These need to be resolved
and eliminated prior to probabilistic assessment.}  External subcase
references need to be expanded in place (like a macro), or separately
evaluated and represented by the confidence assessed for their top
claim (like a subroutine).  Confidence in assumptions is assigned by
assessors or by Subject Matter Experts (SMEs) and is recorded as the
estimated probability that each assumption is true.\footnote{Some of
these estimates may be purely subjective, while others (e.g.,
likelihood of more than $n$ sensor faults) may be based on historical
experience or a combination of reliability analysis, data, and
judgment that should perhaps be expanded into a subargument.}  The
treatment of residual concerns\footnote{Historically, we have spoken
of residual \emph{doubts}, but in a probabilistic context we prefer
residual \emph{concerns} as \emph{doubt} is used for the probabilistic
function $1 - P(a)$.} considers number of occurrences (e.g., are there
tens of minor warnings from a static analyzer or hundreds?)  as well
as likelihood and severity of associated faults, and we postpone its
consideration to Section \ref{residuals}.  Here, we first deal with
confidence in evidence.

\subsection{Confidence in Evidence}
\label{evidence}

\begin{figure}[ht]
\vspace*{-4ex}
\begin{center}
\includegraphics[width=0.7\textwidth]{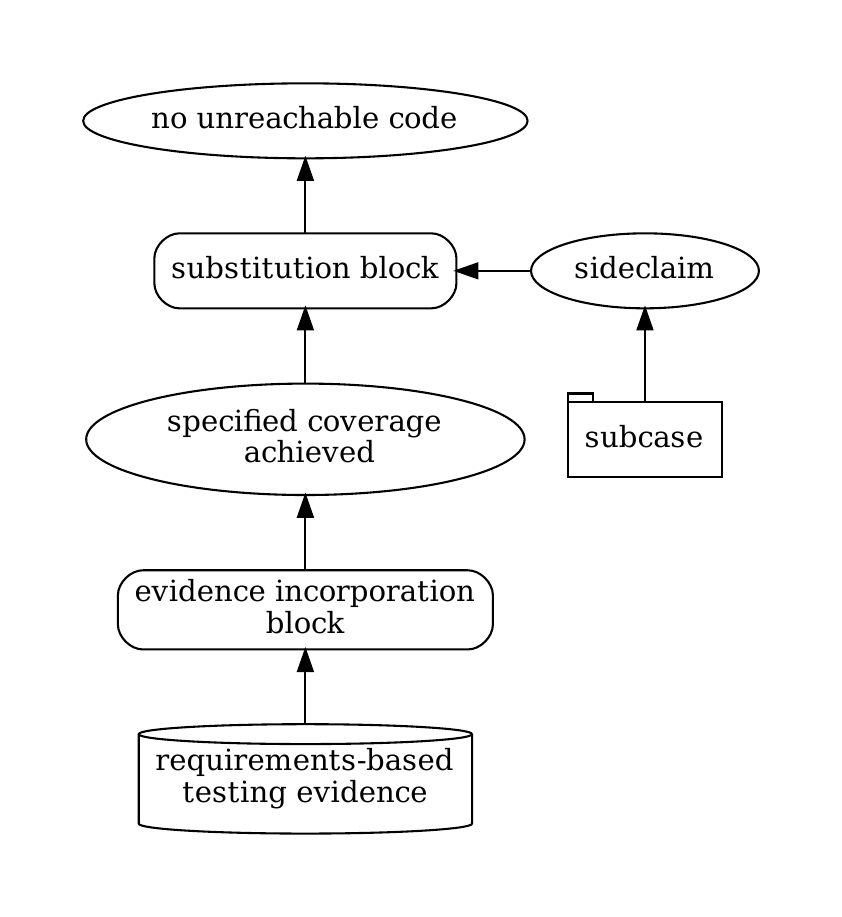}
\end{center}
\vspace*{-6ex}
\caption{\label{evex}Evidence Example}
\vspace*{-2ex}
\end{figure}

Evidence is usually provided to an Assurance 2.0 argument in two
steps, as portrayed in Figure \ref{evex}.  In the first step,
reference to the actual evidence is supplied to an \emph{Evidence
Incorporation} block.  That (building) block or argument step
supports a claim about \emph{something measured}: that is, it tells us
what the evidence \emph{is}.  In Figure \ref{evex} we see an example
where the evidence is provided by requirements-based testing and the
measurement claim states that this achieved some specific level (e.g.,
MC/DC \cite{Hayhurst-etal01}) of structural coverage.  In a full
assurance case, the evidence node and the measured claim will
reference (perhaps via hyperlinks in their labels) detailed accounts
of the testing performed and the coverage obtained, and the evidence
incorporation node will reference a detailed justification that the
evidence indeed supports the claim.  The justification may cite an
established theory for the test protocol involved, with necessary
assumptions referenced in a sideclaim, or it may be decided that
evidence incorporation is premature and that a subargument is needed
to ``make the case'' that the test protocol indeed delivers the
measured claim, thereby elaborating the overall argument beyond that
shown in Figure \ref{evex}.

In the second step, the measured claim is transformed, using a
\emph{substitution} block, into a claim about \emph{something useful}
that tells us what the evidence \emph{means} in the context of this
assurance case.  In the example of Figure \ref{evex} the useful claim
is that the software contains no unreachable code (e.g., debugging
code that is disabled in operation; this is considered hazardous
because experience shows that a fault may cause it to become
reachable, with unpredictable consequences).  The reasoning for this
is documented and justified in the substitution block, citing a
suitable theory of testing and reachability.

In previous papers and reports
\cite{Rushby:AAA13,Bloomfield&Rushby24:CBJ,Bloomfield&Rushby:confidence22}
we describe in detail how claims should be evaluated against evidence
by examining \emph{confirmation measures} constructed from various
estimated conditional probabilities, such as Good's measure
\[\log\frac{P(E \vbar C)}{P(E \vbar \neg\, C)},\] where $C$ represents
the useful evidential claim and $E$ the measured claim.  Confirmation
measures help us assess the discriminating power of our evidence, but
once this has been done and accepted, we need a numerical assessment
of confidence in the useful claim, given the evidence, and for this
the simple posterior probability estimate $P(C \vbar E)$ is
appropriate.  In our notation, this is
\begin{alltt}
   P(useful claim) = P(useful claim\vbar{}measured claim).
\end{alltt}
Note that we are assuming
\texttt{P(measured claim\,\vbar\,evidence)} is 1 because the measured
claim should simply state what the evidence is, but it can be
separately estimated and added as an additional factor if desired.  

Argument blocks in Assurance 2.0 generally have a \emph{sideclaim}
recording conditions that may be necessary to ensure their
justification is sound.  Here, it might specify that coverage must be
measured on executable object code (EOC), not source code.  For
reasons that are explained in the following section, confidence in the
useful claim is then the product of the conditional probability stated
above and confidence in the sideclaim:
\begin{alltt}
   conf(useful claim) =
     P(useful\,claim\vbar{}measured claim) \xx{} conf(sideclaim).
\end{alltt}

The first factor on the right of this assignment must be estimated by
the assessors or other SMEs while the second,
\texttt{conf(sideclaim)}, is evaluated by recursive application of the
techniques being developed here.  

It is also possible to have a sideclaim on the evidence incorporation
block, in which case its confidence will be included as a further
factor in the product above.

\subsection{Confidence in Reasoning Steps}

Above the leaf nodes, an Assurance 2.0 argument consists of reasoning
steps that each justify a parent claim on the basis of one or more
subclaims and, possibly, a sideclaim.  The reasoning steps comprise
concretion, substitution, and exact defeater (aka.\ negation) blocks,
which each have a single subclaim, and decomposition and calculation
blocks that each have two or more subclaims
\cite{Bloomfield&Netkachova14}.

\subsubsection{Blocks with a Single Subclaim}
\label{single}

\begin{wrapfigure}{R}{.5\textwidth}
\vspace*{-9ex}
\begin{center}
\hspace*{-3ex}\includegraphics[width=0.6\textwidth]{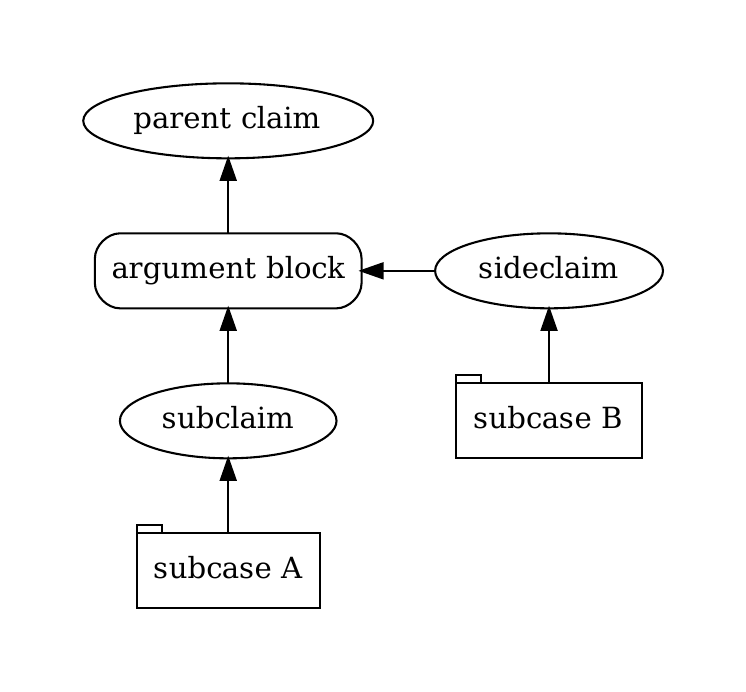}
\end{center}
\vspace*{-8ex}
\caption{\label{singlefig}Argument Block with a\\ Single Subclaim}
\vspace*{-3ex}
\end{wrapfigure}

The substitution blocks that are used in combination with evidence
incorporation blocks have already provided simple examples of argument
blocks with a single subclaim.  The general case for blocks with a
single subclaim is portrayed in Figure\,\ref{singlefig}.

Our task is to estimate probabilistic confidence in the truth of the
parent claim given previously estimated confidence in the components
of its supporting argument block, namely its subclaim, sideclaim, and
the inference used in its argument.

In logical assessment, such as described in
\cite{Bloomfield&Rushby24:CBJ}, we explain that the inference is
assumed to be deductive and the sideclaim states conditions for it to
be sound, which includes the requirement for subclaim(s) to be sound
premises of its argument step.  The logical interpretation of an
argument block is then \texttt{sideclaim \imp (subclaim \imp parent)},
which is equivalent to \texttt{sideclaim \aaa{} subclaim \imp parent}
and so confidence in the parent claim is based on that in
\texttt{sideclaim \aaa{} subclaim}.\footnote{We use \imp for material
implication, and \aaa{} for conjunction.}

We prefer that the argument step is deductive in numerical assessment
also, which can often be achieved by internalizing confidence into
claims (see introduction) and explicitly addressing confidence in its
inference.  Nevertheless, this is not always feasible so we need an
approach that allows some uncertainty and external assessment of
confidence.

Here, confidence in the parent is some function of the argument step
applied to confidence in the subclaim and sideclaim.  That is,
\begin{alltt}
   conf(parent) = f(argstep)(conf(subclaim \aaa sideclaim)).\hfill(1)
\end{alltt}
Where the (higher-order) function \texttt{f} depends on the kind of
reasoning performed in the argstep (e.g., substitution) and its
specific instance (e.g., from structural test coverage to reachable
code).

Confidence is a subjective probability, so we can apply probability
laws.  By the chain rule
\begin{alltt}
   conf(subclaim \aaa sideclaim)
      = conf(subclaim | sideclaim) \xx conf(sideclaim)
\end{alltt}

The first term in the product is confidence in the subclaim under the
conditions stated by the sideclaim, which may seem rather difficult to
assess.  However, we generally expect the subclaim to be conditionally
(though not logically) independent of the sideclaim and so the term
reduces to \texttt{conf(subclaim)} which is assessed by applying the
methods being described here to the subclaim's subcase.

For example, in the previous section the measured claim (i.e.,
subclaim) will state that MC/DC coverage was achieved using some
testing and measurement procedure and confidence in the claim will
concern these procedures and how they were performed.  The sideclaim
will state (possibly among other things) that coverage must be
achieved and measured on EOC\@.  Confidence in this sideclaim will be
1 if the measurement actually was on EOC and 0 otherwise (e.g., if it
was measured on source code), but this has no impact on confidence in
the measured claim.  Other examples may have more graduated confidence
in the sideclaim, but this should generally have no impact on the
subclaim; in cases where it does, the conditional confidence should be
assessed explicitly.  Thus, assuming conditional independence,
\begin{alltt}
   conf(subclaim \aaa sideclaim)
      = conf(subclaim) \xx conf(sideclaim).
\end{alltt}
Substituting back in \texttt{(1)}
\begin{alltt}
   conf(parent) = f(argstep)(conf(subclaim)\,\xx\,conf(sideclaim)).
\end{alltt}

We assert that for the functions \texttt{f} we are interested in, the
product can be decomposed, so that
\begin{alltt}
   conf(parent) \hfill (2)
      = f(argstep)(conf(subclaim))\,\xx\,conf(sideclaim).
\end{alltt}

Applying this formula to Figure \ref{singlefig}, if our confidence in
the subclaim, given its subcase A is 0.8, confidence in the sideclaim,
given its subcase B, is 0.9, and confidence in the inference from test
coverage to reachable code---that is \texttt{f(argstep)(a)}---is
\texttt{0.95 \xx{} a}, then confidence in the parent claim is the
product of these three numbers: 0.684.

\subsubsection{Blocks with Multiple Subclaims}
\label{multiple}

The analysis for argument blocks having multiple subclaims proceeds as
in the previous section, except that subclaim becomes plural in
\texttt{(2)} so that we have
\begin{alltt}
   conf(parent) \hfill (3)
      = f(argstep)(conf(subclaims))\,\xx\,conf(sideclaim).
\end{alltt}

However, the function \texttt{f(argstep)} is now based on two
components: one concerns the inference, as before, and the other
concerns how the multiple subclaims combine to form the premise to
that inference.  We will suppose these elements compose, so that
\begin{alltt}
   conf(parent) \hfill (4)
     =\,f(argstep)(h(argstep)(conf(subclaims)))\,\xx\,conf(sideclaim)
\end{alltt}
where, as before, \texttt{f(argstep)} is the confidence function due
to the inference, as before, and \texttt{h(argstep)} is that due to
the combination of subclaims.  We generally expect blocks to employ
just one or the other of these elements so that one of these functions
will be the identity (or multiplication by a simple factor).  For
example, we often have a substitution block that performs inference
(but no combination) above a decomposition block that performs
combination (but no inference), rather than a single block that
performs both of these.

In the following, we concentrate on combination and will assume
\texttt{f(argstep)} is some simple constant factor \texttt{k} that
defaults to 1, and we will initially postpone consideration of the
sideclaim, so (\hyperref[divform]4) becomes
\begin{alltt}
   conf(parent) = h(argstep)(conf(subclaims)) \hfill (4a)
\end{alltt}
and our task is to estimate \texttt{h(argstep)} for various
kinds of reasoning steps.

We focus on decomposition blocks, which can be used in several
different ways, possibly requiring different probabilistic
interpretations and therefore different calculations for
\texttt{h(argstep)}.

\subsubsection{Decomposition Blocks, Diversity Case}
\label{diversity}

\begin{figure}[ht]
\vspace*{-4ex}
\includegraphics[width=1.0\textwidth]{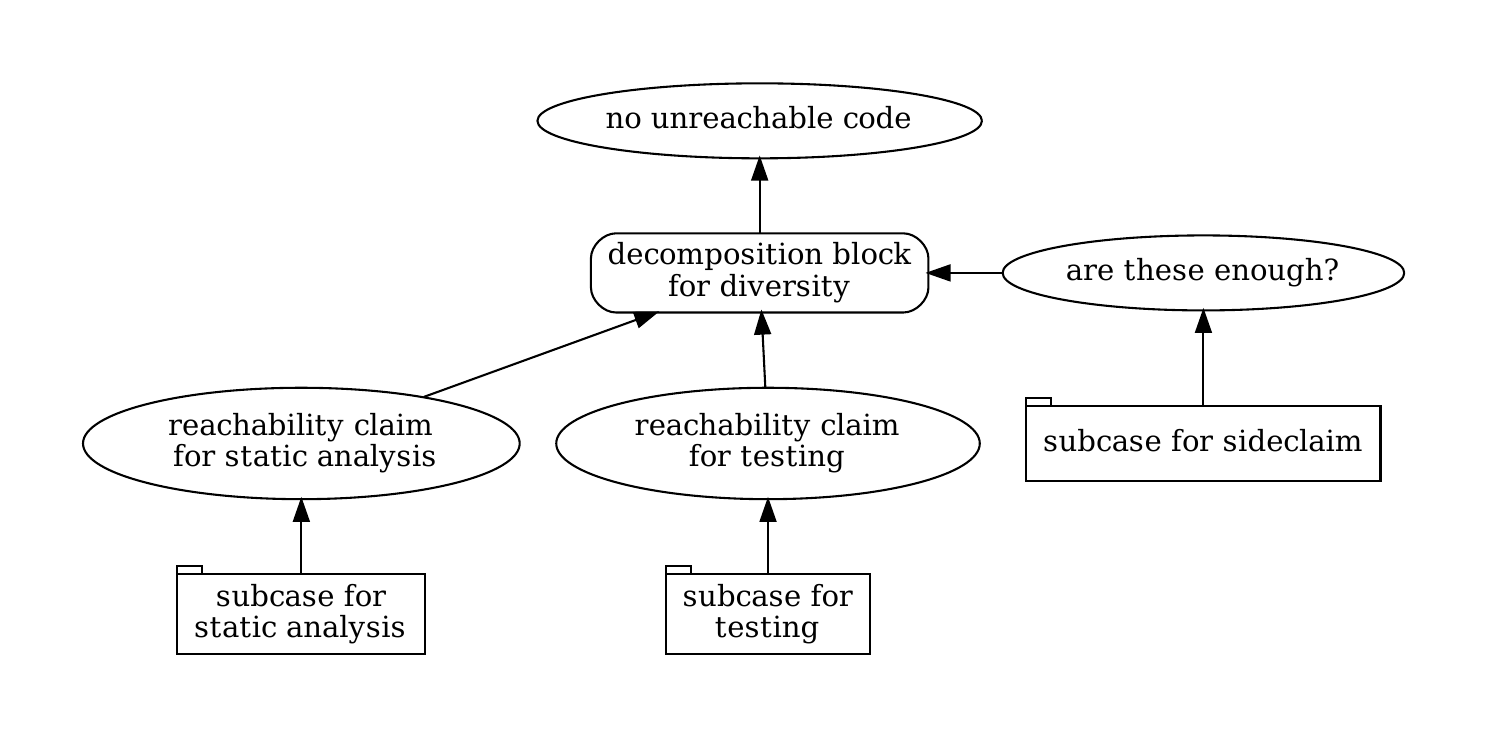}
\vspace*{-7ex}
\caption{\label{diverse}Decomposition Block for Argument by Diversity}
\vspace*{-1ex}
\end{figure}

Let us return to the evidential step of Section \ref{evidence} and
develop it further.  Suppose requirements-based testing did not quite
achieve the desired coverage.  We could try to provide a more complex
justification for the same argument using additional testing, or we
could regard the testing evidence as inadequate and either replace it
or reinforce it with something else.  Testing is attractive because it
provides concrete evidence, so we decide to retain it and buttress it
with static analysis: the idea being that testing and static analysis
build on different foundations and are therefore ``diverse'' and may
be supposed to fail independently \cite{Bloomfield&Littlewood:DSN03}.
This diversity argument is represented in the decomposition block
shown in Figure \nolinebreak \ref{diverse}.

In some discussions of assurance cases, diversity arguments of this
kind are regarded as disjunctions, the idea being that ``if you don't
like one subcase, you can use the other.''  This is not our
interpretation: we think of one subclaim eliminating some \doubts
about the parent claim and the other subclaim eliminating other
\doubts, and we conclude the parent claim only when \emph{both}
subclaims have been assessed as sound.  Hence, the logical
interpretation of a decomposition argument has to be conjunction and
the adequacy of any specific decomposition (i.e., are these specific
subclaims adequate?) is stated as a sideclaim and supported by its own
subcase.

\begin{figure}[ht]
\includegraphics[width=0.5\textwidth]{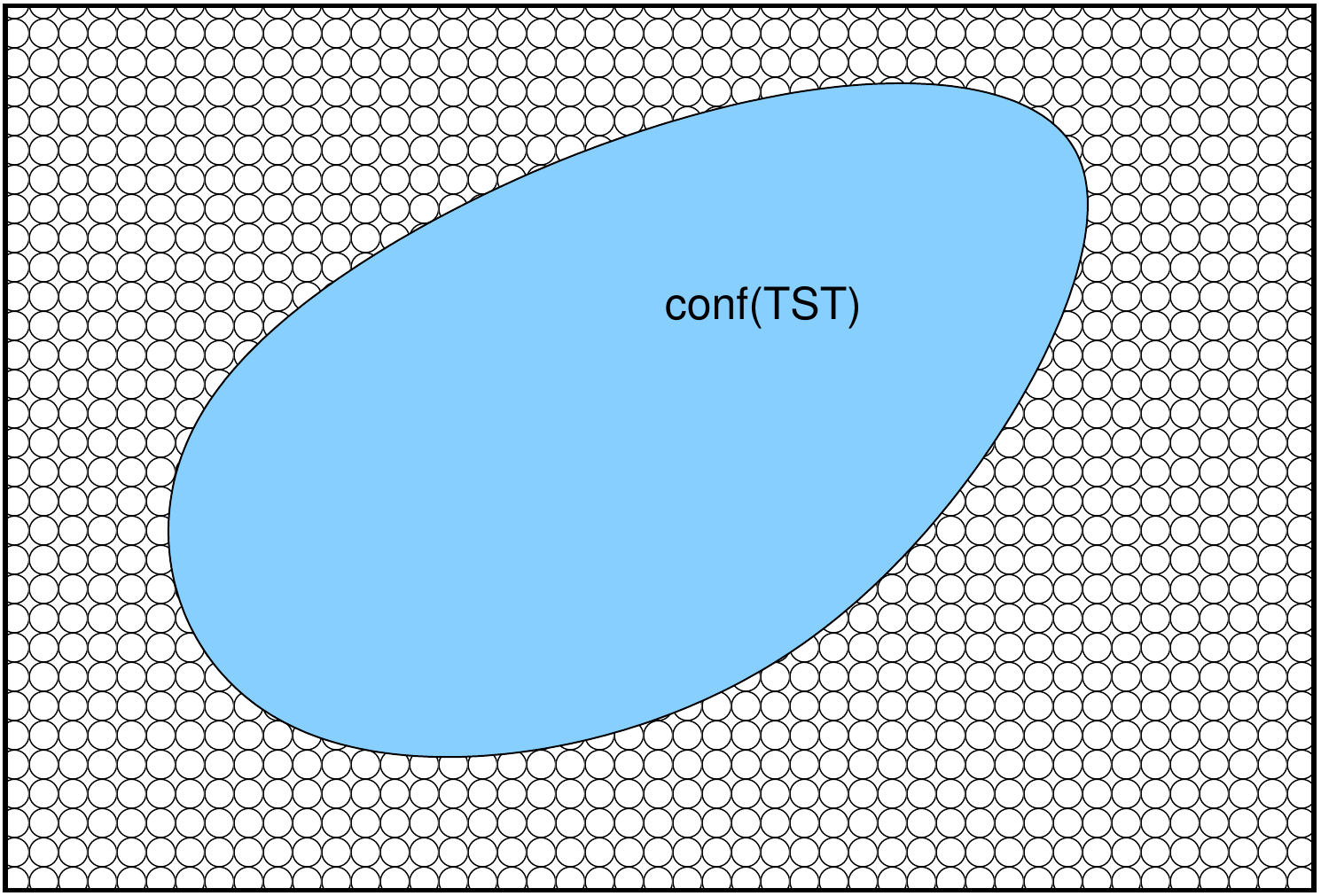}
\includegraphics[width=0.5\textwidth]{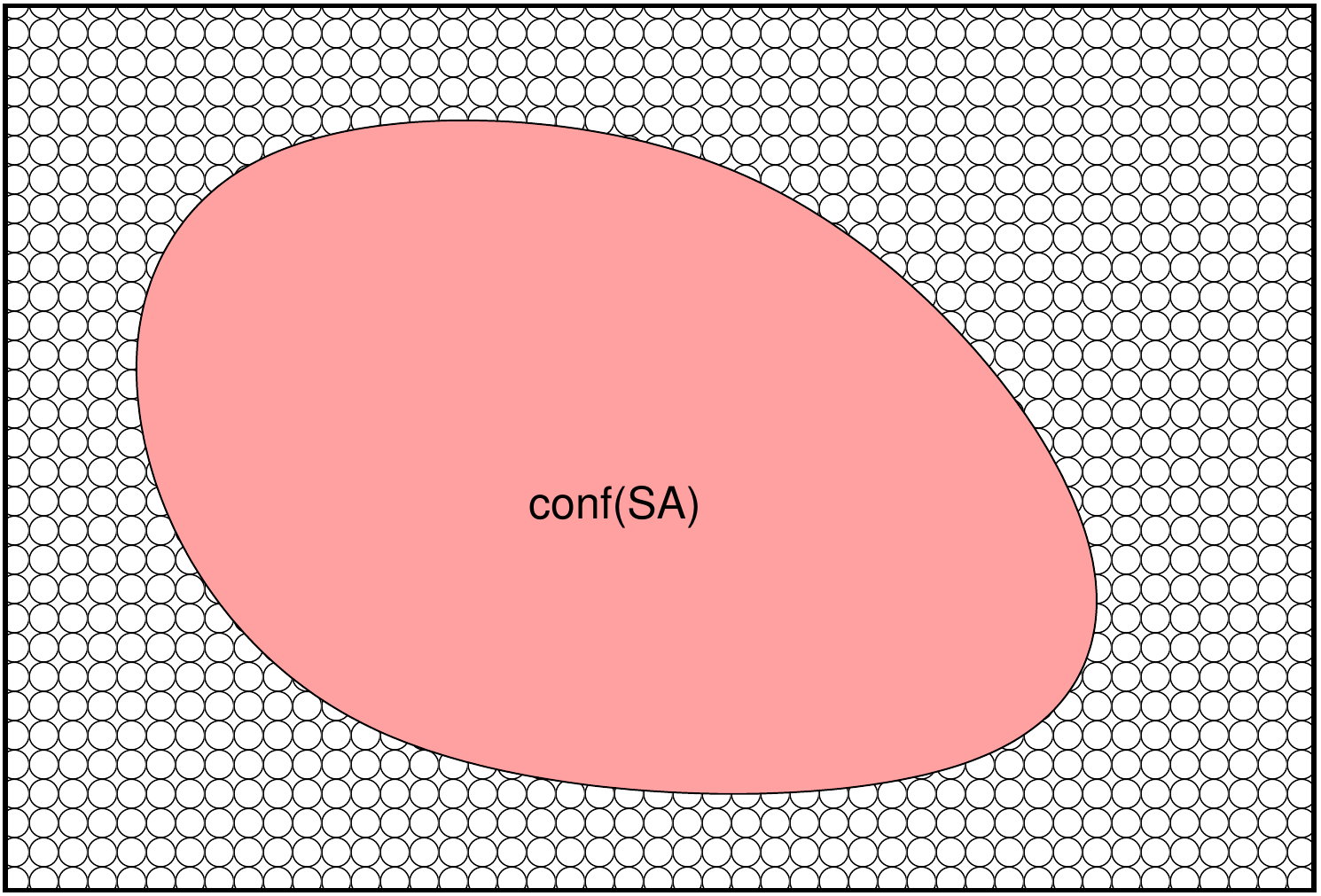}
\vspace*{-2ex}
\caption[\Doubts Eliminated by Testing and by Static
Analysis]{\label{diverseconf1}The background circles represent \doubts
about the parent claim, and the colored shapes represent those
eliminated (i.e., covered) by subclaims for testing (left) and 
static analysis (right)}
\vspace*{-1ex}
\end{figure}

Logical assessment is holistic, but when we assess confidence we are
interested in the \emph{quantity} of \doubts that have been eliminated
and their impact on the parent claim.  This interpretation is
illustrated in the Venn diagrams of Figure \ref{diverseconf1}.  The
blue shape in the left hand diagram indicates the probability mass of
\doubts eliminated by testing and the background circles indicate
those \doubts that remain.  The area of the blue shape represents the
confidence delivered by the testing subcase: \texttt{conf(TST)}.  (Of
course, in practice we want useful evidence to eliminate almost all
\doubts, but we reduce this for pictorial clarity.)  Likewise, we
indicate the \doubts eliminated through static analysis by the pink
shape on the right of Figure \ref{diverseconf1}, where the area of the
shape represents our confidence in the subcase for static analysis:
\texttt{conf(SA)}.

\begin{wrapfigure}{R}{.6\textwidth}
\vspace*{-4ex}
\begin{center}
\includegraphics[width=0.55\textwidth]{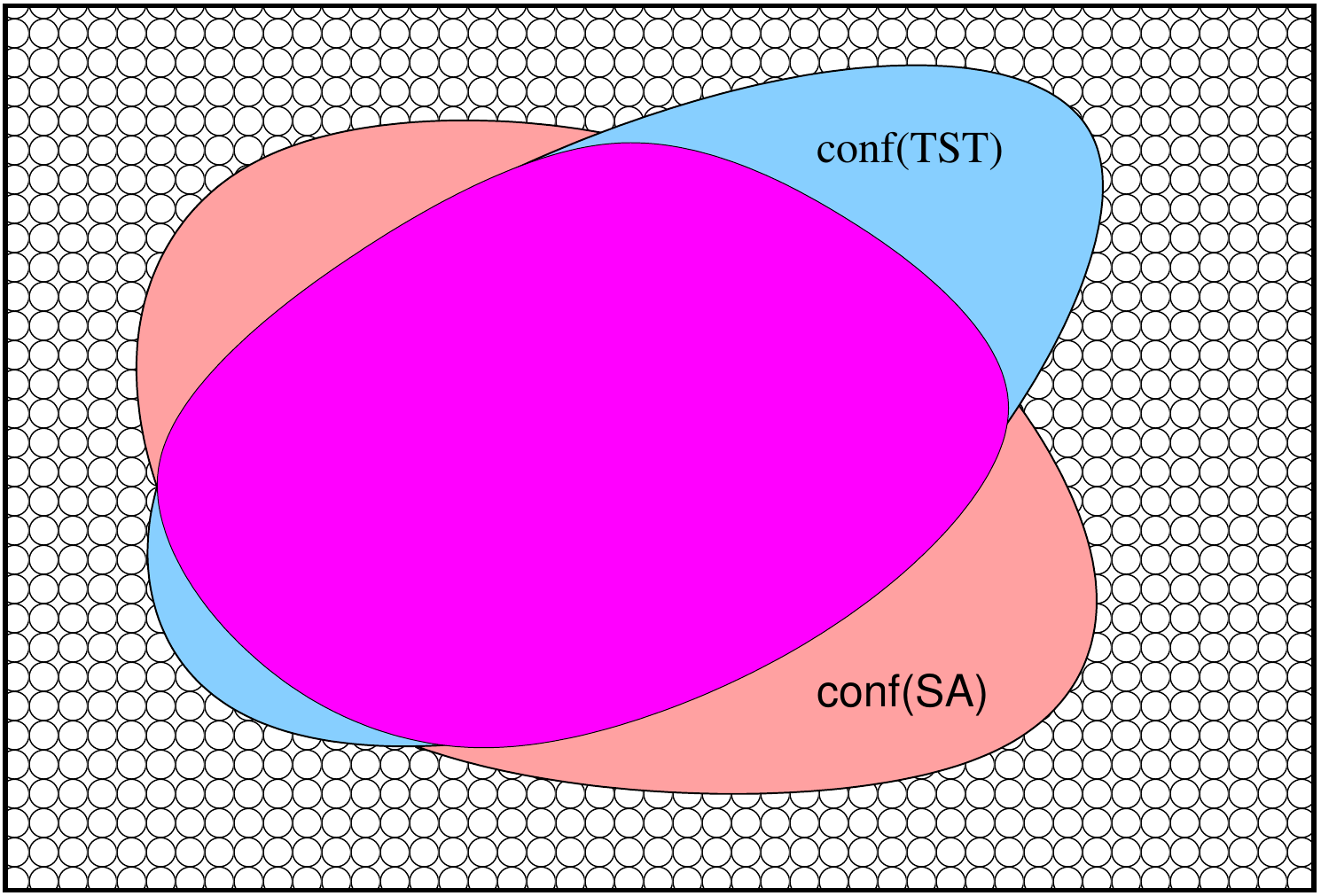}
\end{center}
\vspace*{-2ex}
\caption{\label{diverseconf}\Doubts Eliminated by\\ Combination of Diverse Subclaims}
\vspace*{-3ex}
\end{wrapfigure}

More precisely, we interpret the Venn diagrams over an \emph{epistemic
possibility space}: that is, the (unknown) set of scenarios,
consistent with scope assumptions and background knowledge recorded
elsewhere in the case, that determine whether the parent claim is
\texttt{true} or \texttt{false}.  The intent is epistemic rather than
frequentist: it is a space of possibilities, not a population from
which we sample.

Conceptually, we assign subjective probabilities to scenarios based on
our judgment about their likelihood.  Confidence in a claim
\texttt{C}, denoted \texttt{conf(C)}, is the sum of probabilities
across all scenarios where \texttt{C} is \texttt{true}; 
conversely, its \texttt{doubt} is the sum across
scenarios where it is \texttt{false}.  When we remove $x$\% of doubt,
we gain $x$\% in confidence.

A \emph{\doubt} about claim \texttt{C} is a specific scenario in which
\texttt{C} is \texttt{false}.  In the diagrams, the small background
circles represent such \doubts and the rectangles correspond to the
totality of these (i.e., doubt in the parent claim).  Colored regions
represent sets of scenarios (i.e. events), where the sizes correspond
to the sum of probabilities, the probability mass, associated with
those scenarios.

The \doubts eliminated by the combination of requirements-based
testing and static analysis are shown in Figure \ref{diverseconf}
where their probability mass corresponds to the union of the blue and
pink shapes.  Thus, confidence in the parent claim delivered by the
two subclaims corresponds to the area of this union.\footnote{Notice
how far we have departed from the standard interpretation of
probabilistic logic: \emph{logical} assessment \emph{whether} the
parent claim is true requires the \emph{conjunction} of subclaims,
whereas \emph{confidence} assessment for \emph{how strongly} we should
believe it requires their union (traditionally associated with
\emph{disjunction}).}  This area can be calculated as the sum of the
areas for testing and for static analysis, less their overlap (shown
in magenta), which would otherwise be counted twice.  The overlap
corresponds to the intersection of the testing and static analysis
subclaims.  Thus, using \texttt{NUC}, as an abbreviation for ``no
unreachable code,'' we have\\[-2ex]
\begin{alltt}
   conf(NUC) = conf(SA) + conf(TST) - conf(SA \ii TST)
\end{alltt}
where the right hand side corresponds to
\texttt{h(diversity)(conf(subclaims))} in (4a), and we need some way
to estimate the final term.

We know that static analysis covers a portion of the total ``doubt
space'' equal to \texttt{conf(SA)} so, assuming independence, it also
covers the same proportion of the space covered by testing, which
itself covers \texttt{conf(TST)} of the total space.  Hence, our
estimate for their overlap \texttt{conf(SA \ii\ TST)} is
\texttt{conf(SA)\,\(\times\)\,conf(TST)}\footnote{These and several
other formulas are well-known probabilistic tautologies; we spell them
out so that later derivations are easier to follow.}  and therefore
\begin{alltt}
   conf(NUC) = conf(SA) + conf(TST) - conf(SA)\,\(\times\)\,conf(TST).
\end{alltt}
Recalling that \texttt{doubt(x)} represents \texttt{1 - conf(x)},
this becomes
\begin{alltt}
   conf(NUC)  = 1 - doubt(SA)\,\(\times\)\,doubt(TST), \textrm{and therefore}
   doubt(NUC) = doubt(SA)\,\(\times\)\,doubt(TST).
\end{alltt}
We refer to this calculation as the \emph{product of doubts} and it
generalizes to arguments with $n$ subclaims in the obvious
way:
\setcounter{equation}{4}
\begin{equation}\label{diversityeqn}
\texttt{doubt(parent)} = \prod_{i=1}^n \texttt{doubt(subclaim$_i$)}.
\end{equation}

For notational brevity we have omitted the sideclaim in the
calculations above but we recall from (\hyperref[divform]4) that the confidence that is
propagated is the product of confidence in the inference from
subclaims to parent claim (i.e., the calculation described above) and
confidence in the sideclaim.  Additionally, a function \texttt{f} or
constant factor \texttt{k} can optionally be applied if the developer
or assessor considers it desirable.

So, in the example above, if we are 95\% confident that testing
ensures no unreachable code and 90\% confident that static analysis
does the same, then the doubts are 5\% and 10\% respectively, and
their product is 0.5\%.  Hence, confidence in the combination is
99.5\%.  If we have 90\% confidence in the sideclaim that the
subclaims are adequately diverse and independent, then overall
confidence is the product of these two, which yields
89.5\% (indicating that diversity adds little when there is low
confidence in the sideclaim).

Product of doubts is accurate when the independence assumption is
valid and less so when it is not.  There is more opportunity for
violation of independence when confidence in the individual subclaims
is low.  For example, if we are 49\% confident in each of two
subclaims, the product of doubts calculation gives 74\% confidence in
their combination whereas in reality it could range from 49\% (total
overlap) to 98\% (completely disjoint).  The latter may be a
deliberate, alternative, strategy to diversity, and we consider it
next.

\subsubsection{Decomposition Blocks, Partitioned Case}
\label{parts}

In contrast to the previous example where, in the Venn diagram
representation, the subclaims substantially overlap, we now consider
the case where they are disjoint.  This arises in decompositions over
different hazards, or over the components of a larger system, or over
different operating conditions, etc.  The idea is that the subcases
partition the ``\doubt space'' and the subclaims are each local to a
separate partition.

\begin{figure}[ht]
\vspace*{-4ex}
\begin{center}
\includegraphics[width=1.0\textwidth]{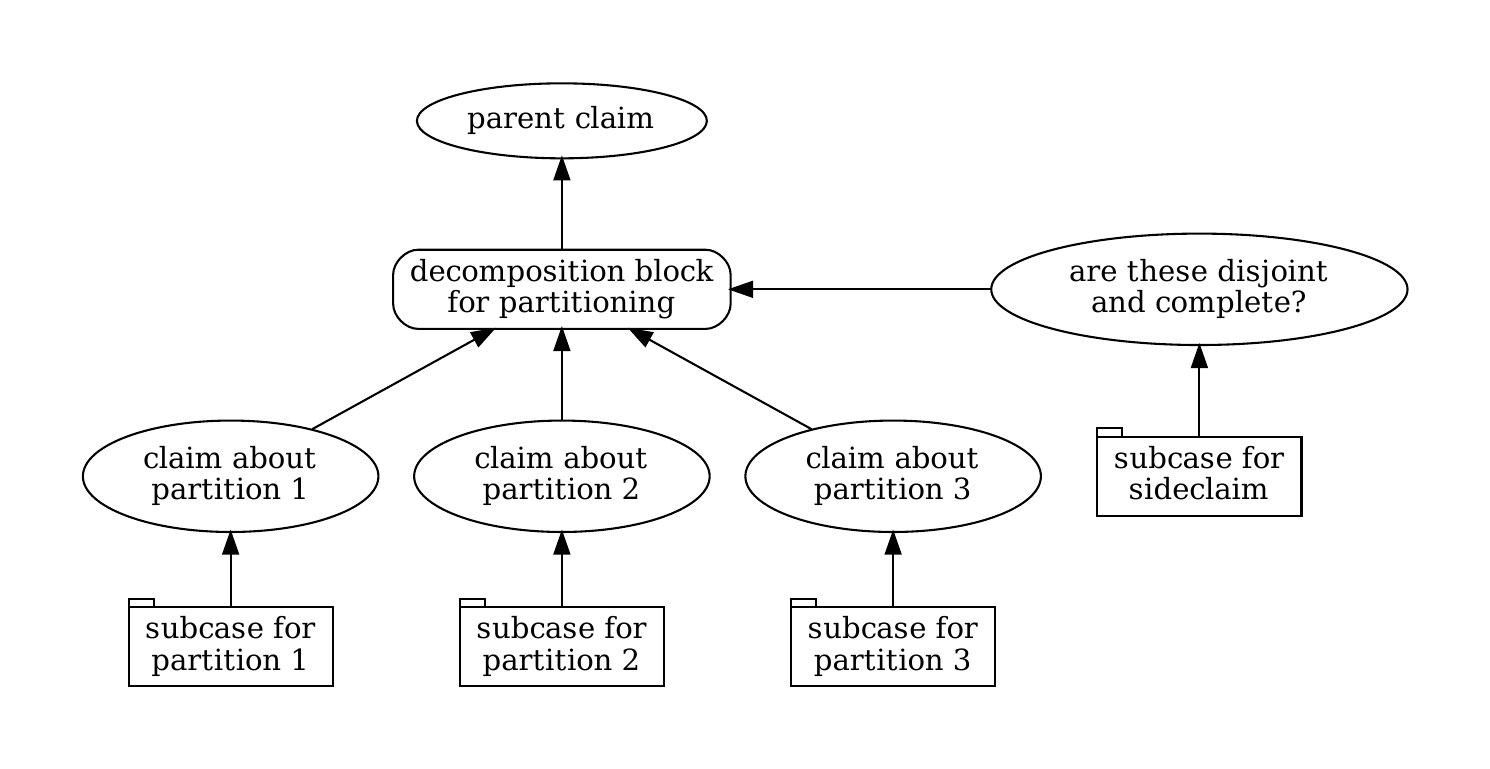}
\end{center}
\vspace*{-6ex}
\caption{\label{partition}Decomposition Block for Partitioned Argument}
\vspace*{-1ex}
\end{figure}

This case is portrayed in Figure \ref{partition}.  To make it
concrete, we could suppose we are decomposing the argument over two
hazards.  This might suggest two partitioned subcases, one for each
hazard, but reflection suggests this is insufficient, we really need
three: one for Hazard 1 alone, another for Hazard 2 alone, and a third
for when they occur together.  It is even conceivable that there
should be further partitions for cases such as those where a second
instance of a hazard arises while the system is still dealing with a
first instance of the same hazard.  However, we will suppose three
partitions are sufficient (e.g., by interpreting the third as
``everything else'') and that this is justified in the sideclaim to
the decomposition.

\begin{figure}[hbt]
\begin{center}
\includegraphics[width=1.0\textwidth]{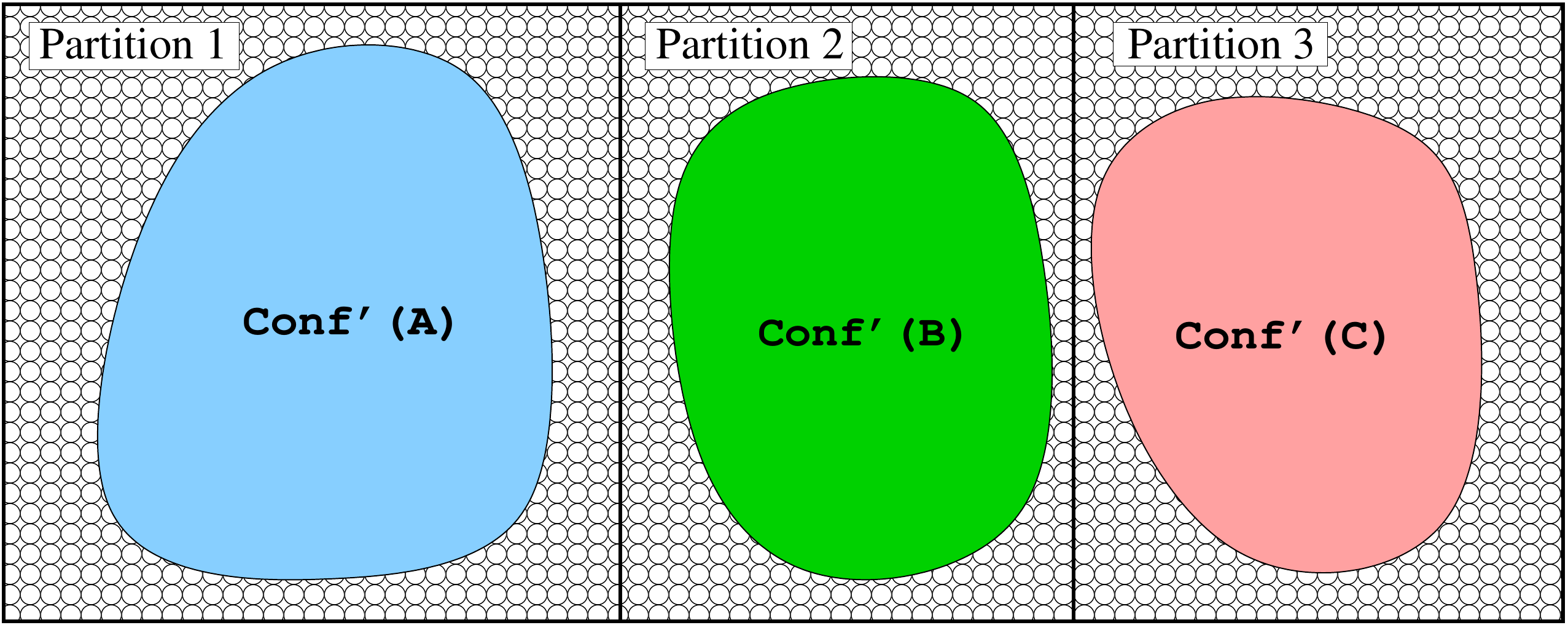}
\end{center}
\vspace*{-2ex}
\caption{\label{partitionconf}Partitioned Subclaims}
\vspace*{-3ex}
\end{figure}

The Venn diagram corresponding to this argument is shown in Figure
\ref{partitionconf} where the background rectangle represents the
total \doubt space partitioned into the three subcases.  The
partitions are of different sizes because we may have more or less
concern about them: \texttt{weight(x)} indicates the fraction of the
total concerns or risks, depending on the claim, associated with
partition \texttt{x} (so the \texttt{weight}s sum to one).  The
colored shapes represent our confidence in the subcase for their
associated partition: \texttt{conf\pp(x)} (we use the prime
\texttt{\pp} to indicate this confidence is assessed relative to its
partition, not the whole space, which is denoted by plain
\texttt{conf(x)} where
\texttt{conf(x)\,=\,conf\pp(x)\(\times\)weight(x)}).  Our confidence
in the overall decomposition, given the sideclaim, is then the
weighted sum of these (this is just a Total Probability calculation):
\begin{alltt}
  conf(parent) = conf\pp(A) \(\times\) weight(1)
                  + conf\pp(B) \(\times\) weight(2)
                  + conf\pp(C) \(\times\) weight(3).
\end{alltt}
Notice that if the \texttt{weight}s are equal, this is simply the
arithmetic mean of the subclaim confidences; we call this the
\textbf{averaging case}.

As in the previous cases, the confidence propagated from this argument
step is the product of confidence in the parent claim, calculated as
above, and confidence in the sideclaim.  Thus if we assess that 60\%
of the risks are associated with Hazard 1 alone, 30\% with Hazard 2
alone, and 10\% with other cases, and that our confidence in the
subclaims associated with these are 90\%, 95\% and 80\%, respectively,
then basic confidence in the parent claim is 90 $\times$ 60 + 95
$\times$ 30 + 80 $\times$ 10 = 90.5\% and this needs to be multiplied
by confidence in the sideclaim, say 99\%, to yield an overall
confidence of 89.6\%.  As in the previous sections, a function
\texttt{f} or constant factor \texttt{k} can optionally be applied to
the basic calculation if the developer or assessor considers it
desirable.

The formula above obviously generalizes to decompositions with $n$
partitioned subclaims as follows.
\begin{equation}\label{parteqn}
\texttt{conf}(\texttt{parent}) =
\sum_{i=1}^n \texttt{conf}^\prime(\texttt{subclaim}_i) \times
\texttt{weight}(\texttt{subclaim}_i)
\end{equation}
where the right hand side corresponds to
\texttt{h(partition)(conf(subclaims))} in (4a).

Despite its different derivation, this is essentially equivalent to
Jeffrey's rule of combination \cite{Ichihashi&Tanaka:Jeffrey89}, which
is (implicitly) used in some Dempster-Shafer treatments of quantified
confidence in assurance cases \cite[Section
3.2]{Ayoub-etal:confidence13}.  Notice that it does not matter if some
of the \doubts addressed by one subclaim are located in the partition
of another, provided assessment of confidence in the subclaim is
confined to its own partition.

Also note that since
\[\texttt{conf}^\prime(\texttt{subclaim}_i) \times \texttt{weight}(\texttt{subclaim}_i)
= \texttt{conf}(\texttt{subclaim}_i),\]
(\ref{parteqn}) is equivalent to
\begin{equation}\label{simplepart}
\texttt{conf}(\texttt{parent}) = \sum_{i=1}^n \texttt{conf}(\texttt{subclaim}_i),
\end{equation}
which corresponds to Fr\'{e}chet's upper bound for union.

We can also derive this result from the analysis used for diversity
cases.  Recall formula \ref{diversity} and imagine pulling apart the
subclaim areas of the \doubt space so that they no longer overlap.
But when there is no overlap, \texttt{conf(A \ii\ B)} is zero.  This
corresponds to the partitioned case and we will obtain the same
results under either the diversity or partitioned calculations,
provided we take note that the subcase confidences are evaluated
differently: in the diversity case they are relative to the whole
\doubt space (i.e., Formula (\ref{simplepart})), whereas in the
partitioned case each is relative to its own partition within that
space and is weighted according to the partition's importance or risk
(i.e., Formula (\ref{parteqn})).

\subsubsection{Decomposition Blocks, Containment Case}
\label{containment}

\begin{figure}[ht]
\vspace*{-1ex}
\includegraphics[width=0.5\textwidth]{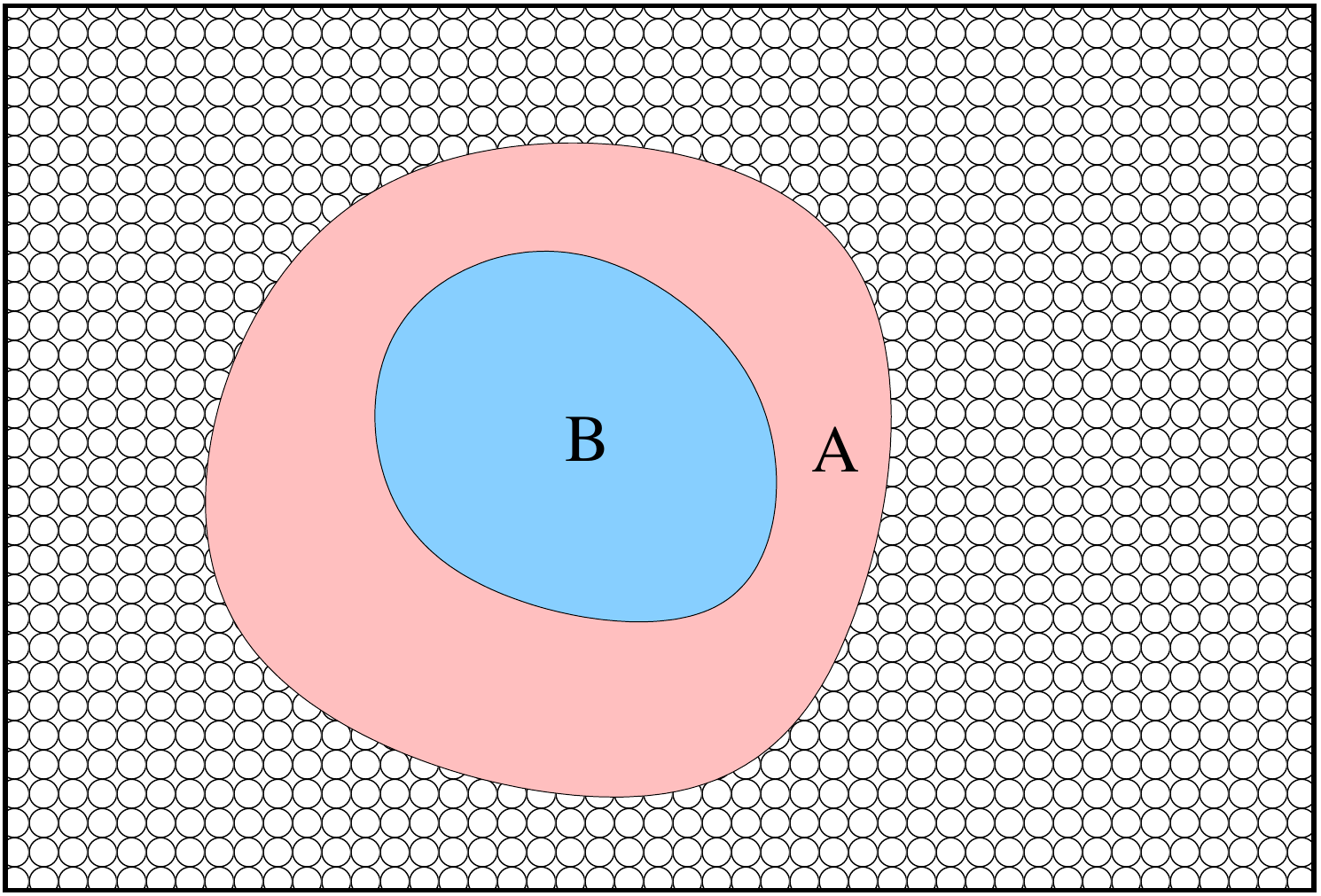}
\includegraphics[width=0.5\textwidth]{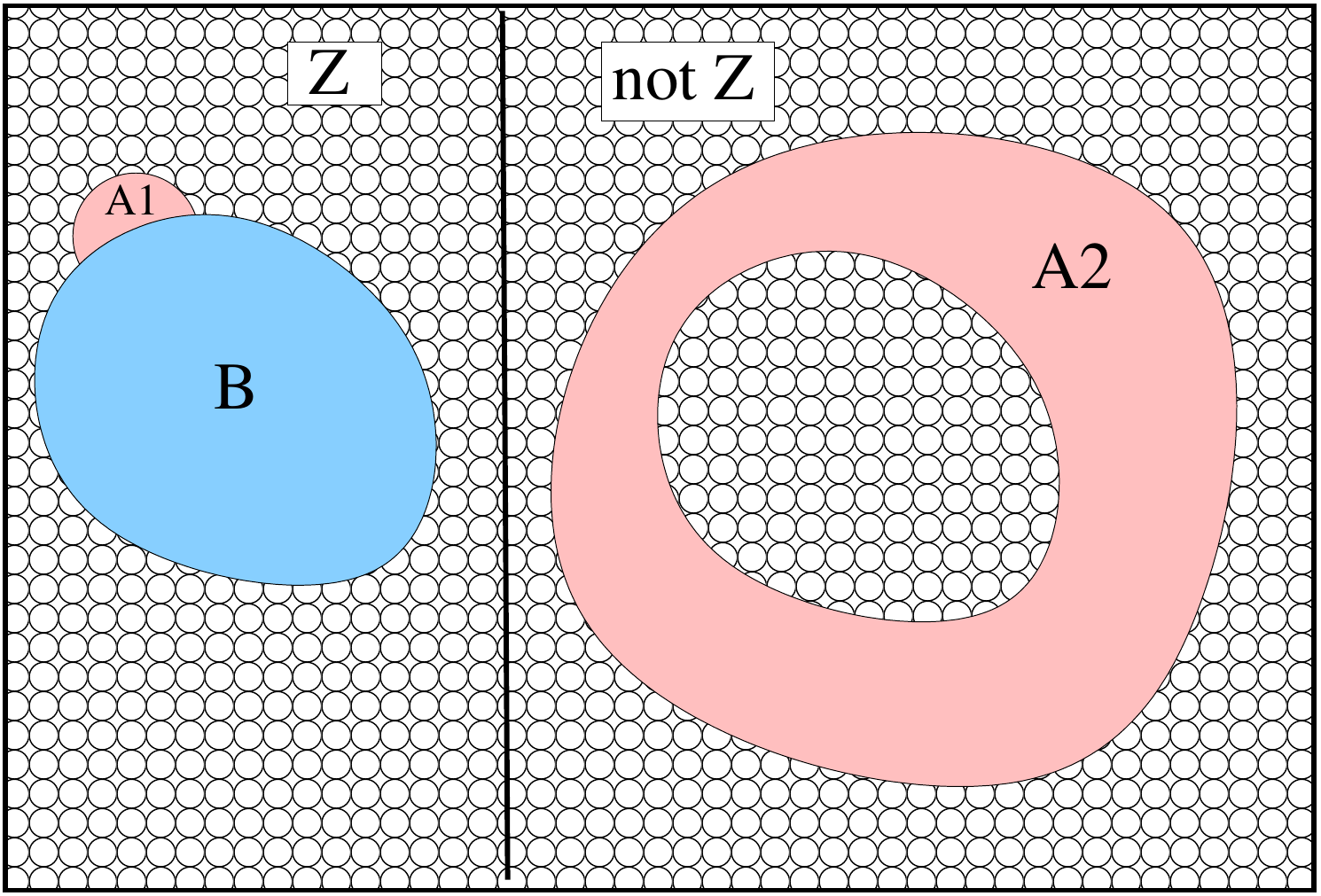}
\caption{\label{contain}Contained Subclaim (on left, redrawn on right)}
\vspace*{-1ex}
\end{figure}

We can reverse the mental experiment just described and imagine
pushing the subclaims of a diversity case together until one is
entirely within the other, as shown on the left of Figure
\ref{contain}.

Ayoub and colleagues \cite{Ayoub-etal:confidence13} call this a
``containment'' argument and treat it as a special case of what they
call a ``disjoint'' argument and that we refer to as a partitioned
case (recall the previous subsection).  We will describe their
treatment later, but first we analyze it as a special form of
diversity case.

It is obvious in the diagram on the left of Figure \ref{contain}
that the eliminated \doubts are all covered by subcase A---and B
adds nothing---so the value propagated should be simply
\texttt{conf(A)}.

Nonetheless, we can attempt to apply our
treatment for diversity cases, as formalized in
(\ref{diversityeqn}).  This gives
\begin{alltt}\label{pform}
   conf(parent) = conf(A) + conf(B) - conf(A) \xx conf(B) \hfill (8)
\end{alltt}
but this treatment is indifferent to how much of B overlaps A: in
particular, it does not propagate the correct value, namely
\texttt{conf(A)}, when B is fully within A and we might suspect it is
also incorrect when only a little of B is outside A\@.

Now (\hyperref[pform]{8}) is derived from our treatment of diversity
cases in Section \ref{diversity}, which starts from the observation,
true in all cases, stated in (\hyperref[divform]{4}).  The challenge
in applying (\hyperref[divform]{4}) is to find a good estimate for
\texttt{conf(A \ii\ B)}.  The treatment in (\hyperref[pform]{8})
assumes that A and B are independent, so that \texttt{conf(A) \xx
conf(B)} is a good estimate for \texttt{conf(A \ii\ B)}.  However, in
the case where B is within A, the subcases are positively correlated
and that product is no longer a good estimate.  But a conservative
estimate is provided by the \emph{Fr\'{e}chet upper bound for
intersection} \cite{frechet-wiki}:
\begin{alltt}
   conf(A \ii B) \(\leq\) \(\min\)(conf(A), conf(B)).
\end{alltt}
In our containment case, the minimum is \texttt{conf(B)} and applying
this in (\hyperref[divform]4) gives \texttt{conf(parent) \(\geq\) conf(A)} as required.
More directly, (\hyperref[divform]4) becomes
\begin{alltt}
   conf(parent) \(\geq\) \(\max\)(conf(A), conf(B)),
\end{alltt}
which corresponds to the \emph{Fr\'{e}chet lower bound for
union}.  For the general case of containment among $n$
subclaims, we have
\[\texttt{conf}(\texttt{parent}) \geq \max_{i=1..n}(\texttt{conf}(\texttt{subclaim}_i)),\]
which comes from the general form of the Fr\'{e}chet bound
\cite{frechet-wiki}.

For completeness, we recall the \emph{Fr\'{e}chet lower bound for
intersection}:
\begin{alltt}
   conf(A \ii B) \(\geq\) \(\max\)(0, conf(A) + conf(B) - 1).
\end{alltt}
When \texttt{conf(A)} and \texttt{conf(B)} are fairly large (i.e., sum
to more than 1), substituting into (\hyperref[divform]{4}) gives
\texttt{conf(parent) \(\leq\) 1}, which is not very helpful, but when
their sum is less than 1 (as will likely be so with largely disjoint
subcases), it gives
\begin{alltt}
   conf(parent) \(\leq\) conf(A) + conf(B),
\end{alltt}
which is an alternative way of deriving the basis for our treatment of
partitioned cases in (\ref{simplepart}).

We see that our general treatment of diversity cases is sound but the
specifics of its application depend on whether the subcases are
independent or not.  If they are independent, the product of doubts
calculation provides a good estimate; if they are positively
correlated (i.e., largely nested) then a calculation using the
Fr\'{e}chet lower bound for union provides a conservative estimate,
while if they are negatively correlated (i.e., largely disjoint) the
partitioned treatment (which corresponds to Fr\'{e}chet's upper bound
for union) is suitable.  In practice, we may not know if our subcases
are correlated, so we can do ``what if'' exercises to examine impact
of the alternative possibilities, with further examination should it
prove significant.

As mentioned earlier, Ayoub and colleagues
\cite{Ayoub-etal:confidence13} treat containment cases as a special
form of disjoint cases, which are what we call partitioned cases.  We
next describe this treatment as it provides an alternative (but more
contrived) route to the same conclusion.

The idea of their construction is to create two partitions: one
corresponding to B and the other to everything else.  We adjust this
slightly, as illustrated by the diagram on the right of Figure
\ref{contain}, so that \doubts covered by B are contained in a
partition Z, and \doubts covered by A are divided into two parts: A1
contains the \doubts also covered by B (we show a little of A1 for
clarity but it is really hidden under B), and A2 contains
all the other \doubts covered by A and is located in partition
``\texttt{not Z}.''  By formula (\ref{parteqn}), we
have
\begin{alltt}
   conf(parent) = conf\pp(B)\xx{}weight(Z) + conf\pp(A2)\xx{}weight(not Z).
\end{alltt}
Now, \texttt{conf\pp(B) = conf(B) / weight(Z)},\\ and
 \texttt{conf\pp(A2) = conf(A2) / weight(not Z)}, so
\begin{alltt}
   conf(parent) = conf(B) + conf(A2).
\end{alltt}
But
 \texttt{conf(A2) = conf(A) - conf(A1)} and
 \texttt{conf(A1) = conf(B)}, and thus
\texttt{conf(parent) = conf(A)} as desired.

This is the same (correct) result that we obtained previously using a
diversity case adjusted to use Fr\'{e}chet's upper bound for union.
This is not surprising as all our calculations are variants on the
same basic method, but it is gratifying to see that different routes
through the constructions do converge on the same result.

Discussion of these constructions does invite the question why should
we be interested in containment cases?  We examined them because Ayoub
and colleagues do but otherwise see no utility in them.  They do,
however, provide a conservative limiting case for diversity arguments
that may be useful when unable to justify the independence assumptions
needed for stronger results.

\subsubsection{Decomposition Blocks, Cumulative Case}
\label{cumulative}

Barrett and colleagues \cite[page 32]{Barrett-etal:confidence25}
consider a case where the parent claim that a dangerous AI function is
taken out of service within one week is supported by the following
three subclaims:
\begin{description}

\item[Detection:] Incident monitoring detects all novel cyber attacks
within one day,

\item[Revision:] A revision protocol ensures the safety case is
 updated within five days of a novel cyber attack being detected,

\item[Removal:] The AI system will be taken offline within one day of
 the top-level safety case claim becoming false.

\end{description}

They represented the argument by a decomposition block similar to
Figure \ref{partition} and estimated confidence in the detection,
revision, and removal subcases to be 48\%, 81\%, and 56\%
respectively.

At the time they performed their study, the only probabilistic
assessments offered for Assurance 2.0 were the ``product'' and ``sum
of doubts'' calculations outlined in Section \ref{oldmethod}.  The sum
of doubts is conservative but robust (recall, it is equivalent to the
Fr\'{e}chet lower bound), while the product calculation has rather
strong assumptions.  Here, sum of doubts delivers confidence 0 in the
parent claim, while the product calculation delivers 21.772\%.

Now that we have developed more specific treatments it is worth
reexamining these calculations to see if we can derive a more
precise result.  However, although the argument resembles Figure
\ref{partition}, we can see that the subcases are not independent
(each is contingent on the success of its predecessors) so the
partitioned treatment is not appropriate, and neither is diversity.

\begin{wrapfigure}{R}{.5\textwidth}
\vspace*{-2.5ex}
\begin{center}
\includegraphics[width=0.5\textwidth]{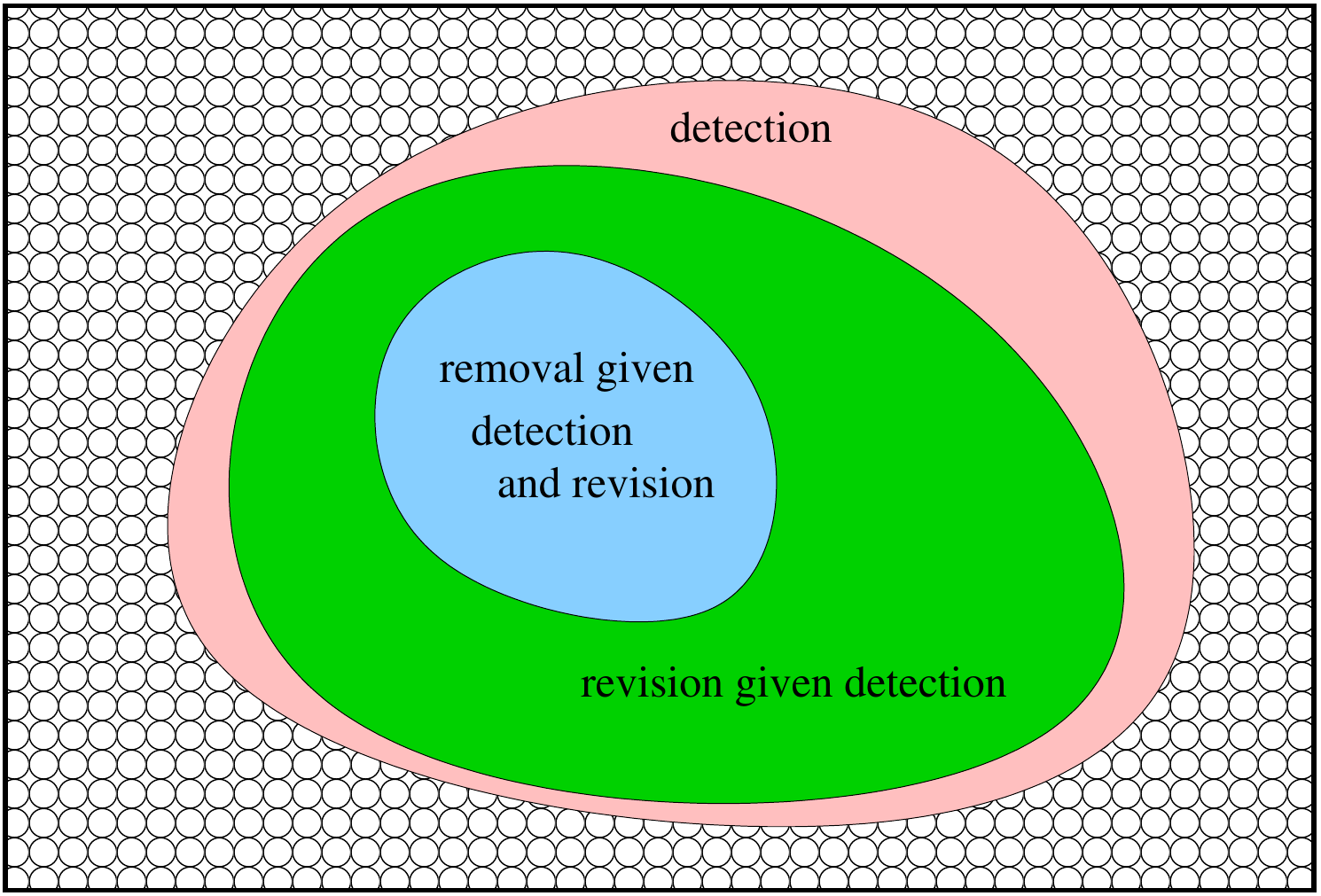}
\end{center}
\vspace*{-3ex}
\caption{\label{conjconf}Cumulative Subclaims}
\vspace*{-2ex}
\end{wrapfigure}

If we try to portray the \doubts eliminated by each subcase, we obtain
Figure \ref{conjconf}, which looks like Figure \ref{contain} but we
can see that it requires a different interpretation.  Rather than the
\doubts eliminated by later subclaims being contained within those of
earlier ones, they build support \emph{cumulatively} and are all
required.  That is, by the standard probabilistic interpretation of
conjunction (the chain rule):
\begin{alltt}
   conf(parent) = conf(detection \& revision \& removal)
                = conf(detection)
                  \xx conf(revision\vbar{}detection)
                  \xx conf(removal\vbar{}detection \& revision).
\end{alltt}

Now, notice that the definitions of the \texttt{detection} and
\texttt{removal} subclaims given earlier implicitly have this
conditional form.  Hence, the estimated subclaim confidences do
correspond to the terms in the formula above and this is the same as
the product calculation performed earlier---but now we can see that
for this case it is exact, rather than an approximation.

This example is a simple illustration of an argument where there are
conditional relationships among subclaims.  Because the relationships
in this example are simple, we can easily calculate (as the product)
the confidence to be propagated to the parent claim (subject, as
usual, to adjustment by a function \texttt{f} or factor \texttt{k} and
multiplication by confidence in any sideclaim).  In arguments where
the conditional relationships among subclaims are more complex than
illustrated here, it may be useful to employ BBNs and their tools,
which are discussed in Section \ref{bbns}.

\begin{table}[phtb]

\centering
\begin{tabular}{|p{1.70in}|p{1.90in}|p{0.80in}|}

\hline \textbf{Approach} & \textbf{Confidence Formula} &
\textbf{Example\newline Value}\\
\hline\hline

Sum of doubts (Fr\'{e}chet intersection lower bound) & $1 - \min(1, d_1 +
d_2 + d_3)$ or\newline $\max(0, c_1+c_2+c_3-(3-1)) $ & 0.55\\\hline

\multicolumn{3}{|l|}{Most conservative; no dependency assumptions} \\
\hline\hline

Confidence Product\newline & $c_1 \times c_2 \times
c_3$ & 0.612 \\\hline

\multicolumn{3}{|l|}{Assumes independent subclaims} \\

\hline\hline

Containment (Fr\'{e}chet\newline union lower bound)  & $\max(c_1, c_2,
c_3)$ & 0.9\\\hline

\multicolumn{3}{|l|}{Total dependence (perfect positive
correlation among subclaims)} \\

\hline\hline
Diversity\newline (Product of Doubts) & $1 - d_1 \times d_2 \times d_3$ & 0.997\\\hline

\multicolumn{3}{|l|}{Strong assumption: doubts are independent} \\

\hline\hline

Arithmetic Mean\newline (Averaging Case) & $1/n \times\sum c_i^{\,\prime}$
& 0.85\\\hline

\multicolumn{3}{|l|}{Equally weighted disjoint partitions (perfect
negative correlation)} \\

\hline\hline
Weighted Average\newline (Partitioned Case) & $\sum w_i \times
c_i^{\,\prime}$; e.g.,\newline
$0.5 \times 0.9 + 0.3 \times 0.8 + 0.2 \times 0.85$ & 0.865\\\hline

\multicolumn{3}{|l|}{Weighted disjoint partitions} \\

\hline\hline

Conditional Chain\newline (Cumulative Case) & $c_1 \times P(c_2 | c_1) \times P(c_3 | c_1 \wedge c_2)$ & 0.761\\\hline

\multicolumn{3}{|l|}{Used when subclaims build on each other in sequence} \\

\hline
\end{tabular}

\caption[Summary of Confidence Formulas]{\label{table} Summary of
formulas for several approaches to probabilistic confidence
assignments in Assurance 2.0.  The example is a decomposition block
with three subclaims having confidence $c_1, c_2, c_3$ with values
0.9, 0.85, 0.8, and conditional values $P(c_2 | c_1) = 0.89$ and
$P(c_3 | c_1 \wedge c_2) = 0.95$.  Doubts ($1-c_i$) are denoted $d_1,
d_2, d_3$.  Confidence within a partition $c_i \div w_i$ is denoted
$c_i^{\,\prime}$ and the example values are reinterpreted as those
$c_i^{\,\prime}$.}

\end{table}

\subsubsection{Decomposition Blocks, Other Cases}
\label{others}

In Sections \ref{diversity} (diversity), \ref{parts} (partitioned),
and \ref{containment} (containment) we presented a repertoire of
different probabilistic interpretations for the logical conjunction in
a decomposition block.  Furthermore, we have seen that partitioned and
cumulative cases can be reduced to diversity cases, so that all are
based on a single construction that also clarifies analysis of
cumulative cases.  We consider this uniformity to be a positive
indication that our treatment is correct and we speculate that similar
constructions can be developed for other novel cases.

Table \ref{table} summarizes the various probabilistic formulas that
we have introduced, along with calculated values for a simple example.

\subsection{Residual Risks}
\label{residuals}

An assurance argument may contain residual \doubts (also called
residual doubts, recall footnote \ref{resids}): these are explicitly
marked defeaters that we have been unable or have chosen not to
eliminate or fully mitigate.

Residual \doubts may be due to aleatoric uncertainty (i.e.,
uncertainty in the environment): for example, the system may be
designed to withstand a single sensor failure and historical evidence
indicates this is sufficient, but it is always possible to encounter
more.  Or they may be due to epistemic uncertainty such as the
fallibility of human review (e.g., human requirements tracing cannot
be guaranteed to be free of error), or to limitations in automated
analysis (e.g., automated static analysis may be unable to discharge
some proof obligations, leading to possible false alarms that must be
reviewed by humans, a potentially error-prone process).  More
generally, they may represent a missing or unknown subcase in a
nondeductive (i.e., ``inductive'') reasoning step.\footnote{Residual
\doubts may also be used to represent nonspecific \doubts that we are
unable to localize more specifically.}

In assessing logical soundness in an assurance case, we assume the
consequences of residual \doubts (i.e., residual risks) are
insignificant and, on that basis, we ignore them.  We thereby incur an
obligation to justify this assumption and must consider the potential
impact that a faulty assessment could have on failure (i.e.,
defeasibility) of the case.  Specifically, for each residual \doubt,
we must show that the likelihood of wrongly assessing it (as
residual), combined with its worst possible consequences (i.e., its
\emph{risk}), is below some threshold for concern.

We usually assess this threshold in a qualitative manner, as described
in the fourfold classification below, which is adapted from
\cite{Bloomfield&Rushby24:CBJ}.
\begin{description}

\item[Significant:] an individual residual \doubt poses a risk that is
    above the threshold for concern.  In this case, the issue cannot
    be considered a merely ``residual'' risk, but must treated as a
    defeater and either eliminated or mitigated.

\item[Minor:] an individual residual \doubt poses a risk that is below
    the threshold for concern, but it is possible that many such might
    cumulatively exceed the threshold.  An example could be static
    analysis, where we use human review to evaluate proof obligations
    that the automation cannot decide.  These risks need to be managed
    explicitly: 10 might be OK, but not 100.

\item[Manageable:] a class of minor residual risks whose number and
cumulative severity are below the threshold of concern.

\item[Negligible:] these are residual \doubts where even multiple
    similar instances collectively pose a risk that is below the
    threshold for concern.  This may arise when a source of \doubt
    occurs many times but is adjudged to be trivial.  An example
    (depending on local policy) might be ``style'' warnings from a
    static analyzer.

\end{description}
At final assessment, the only residual \doubts remaining should be
those whose risks are categorized negligible and those categorized
minor but manageable.

In probabilistic assessment, we can follow the lead of logical
assessment and ignore residual \doubts on the basis that the
classification above ensures they are truly negligible.
Alternatively, we can attach a numerical assessment to each residual
\doubt and propagate this through the argument in the usual way.  The
numerical assessment is our subjective confidence that the residual
\doubt will have no impact and should be justified in an accompanying
narrative that considers the likelihood of harm due to the identified
\doubt, its maximum severity, and number of instances if it is a
circumstance that may occur many times.

\section{Bayesian Belief Networks}
\label{bbns}

As noted in the Introduction, we have covered this topic before
\cite{Rushby:AAA15}, but the presentation here is integrated into our
larger narrative.  In particular, it builds on Section
\ref{cumulative} to show how confidence can be calculated for argument
steps with complex dependencies among its subclaims.

The example in Section \ref{cumulative} has three subclaims that
depend cumulatively on each other; this dependency is sufficiently
straightforward that we can calculate confidence in their combination
as a simple function (i.e., the product) of confidence in each of them
individually.  Here, we examine an example with more complicated
dependencies and illustrate how tools for Bayesian Belief Networks can
be used to perform the required calculations.

\begin{figure}[ht]
\vspace*{-4ex}
\begin{center}
\includegraphics[width=1.0\textwidth]{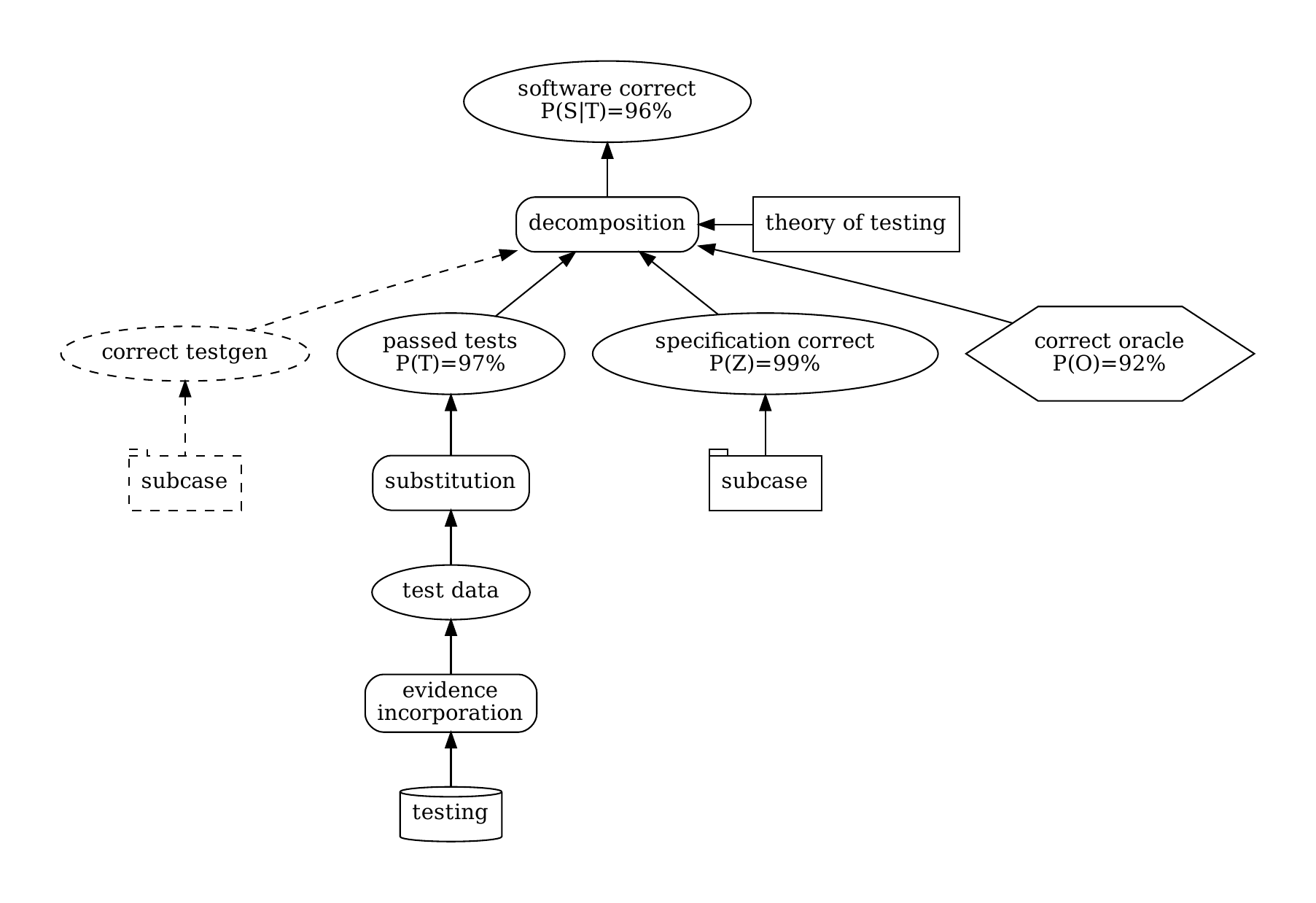}
\end{center}
\vspace*{-6ex}
\caption{\label{bbn-eg}Argument for Correctness by Testing}
\vspace*{-1ex}
\end{figure}

The example concerns assurance by testing for software correctness.
The claim of correctness by testing depends on the actual correctness
of the software (wrt.\ its specification), correctness of the
specification (wrt.\ its informal requirements), correctness of the
test oracle (wrt.\ the specification), coverage of the test generation
process, a suitable theory of testing, and the outcome of the testing
process.  A skeletal assurance argument for this example is shown in
Figure \ref{bbn-eg}.  The hexagonal node shape for \texttt{correct
oracle} indicates that this is an assumption: there is no supporting
evidence.  To keep the presentation succinct we will ignore the test
generation process (it would not illustrate any new topics), so its
subcase is shown with dashed lines.

The node labels of Figure \ref{bbn-v2} include confidence measures,
where that for the decomposition block that delivers the top claim is
calculated using the averaging method of Section \ref{parts} (and
those for its subclaims are simply assumed).  This calculation assumes
that the subclaims to the decomposition are independent, whereas in
reality the software passing its tests depends on its own correctness,
the correctness of the test oracle, and the quality of the testing
process (wrt.\ the theory employed).  Correctness of software that has
passed the tests then depends on correctness of its specification.
These are fairly complex dependencies that are not represented in the
Venn diagrams used in Section \ref{parts}.  Instead, we turn to
Bayes' Theorem.

Bayes' Theorem is the principal method for analyzing conditional
subjective probabilities or beliefs: it allows a prior assessment of
probability to be updated by new evidence to yield a justified
posterior probability.  It is difficult to calculate over complex
conditional (i.e., interdependent) probabilities, but usually the
dependencies are relatively sparse and can conveniently be shown by a
graph or ``network,'' giving rise to the term ``Bayesian Belief
Network'' or BBN\@.  Software tools based on Bayes' Theorem are able
to exploit the sparseness and can calculate probabilities associated
with various scenarios relevant to the problem modeled by a BBN\@.

\begin{figure}[ht]
\begin{minipage}[t]{2.9in}
\vspace*{0.05in}
\includegraphics[width=2.5in]{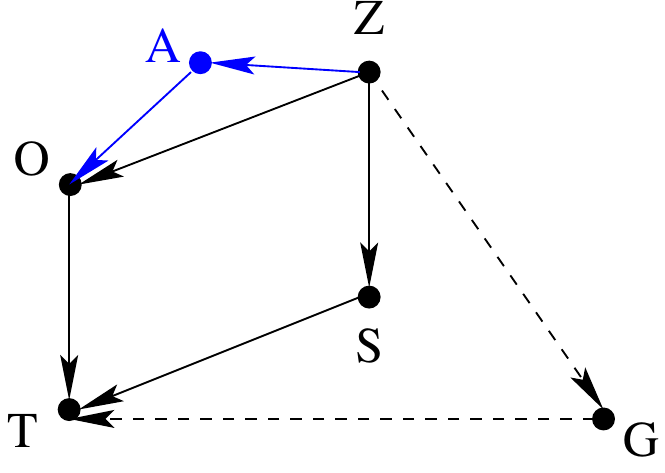}
\end{minipage}\hfill\begin{minipage}[t]{2.3in}\raggedright
\vspace*{0.03in}
\begin{description}\itemsep=1ex
\item[\textbf{Z}:] System Specification
\blue{\item[\textbf{A}:] Specification quality}
\item[\textbf{O}:] Test Oracle
\item[\textbf{S}:] System's true quality
\item[\textbf{T}:] Test results
\item[\textbf{G}:] Test generator
\end{description}
\end{minipage}
\caption{\label{bbneg}BBN for Testing Evidence}
\end{figure}

A BBN for our testing example, adapted from
\cite{Littlewood&Wright:tse07} and \cite{Rushby:AAA15}, is shown in
Figure \ref{bbneg}.  The nodes of the graph represent probabilistic
judgments about components of the argument and the arcs indicate
dependencies between these.

More precisely, the nodes of the graph represent random variables but
we can most easily understand the construction of the graph by first
considering the artifacts from which these are derived.  Here, $Z$ is
the system specification; from this are derived the actual system $S$,
the test oracle $O$ (ignore, for the time being, the arcs associated
with $A$ and shown in blue) and the test generator $G$.  Tests $T$ are
dependent on the test generator, the oracle, and the system.  As
before, we will ignore test generation and these arcs are shown as
dashed lines.

\begin{figure}[ht]
\begin{center}
\includegraphics[width=\textwidth]{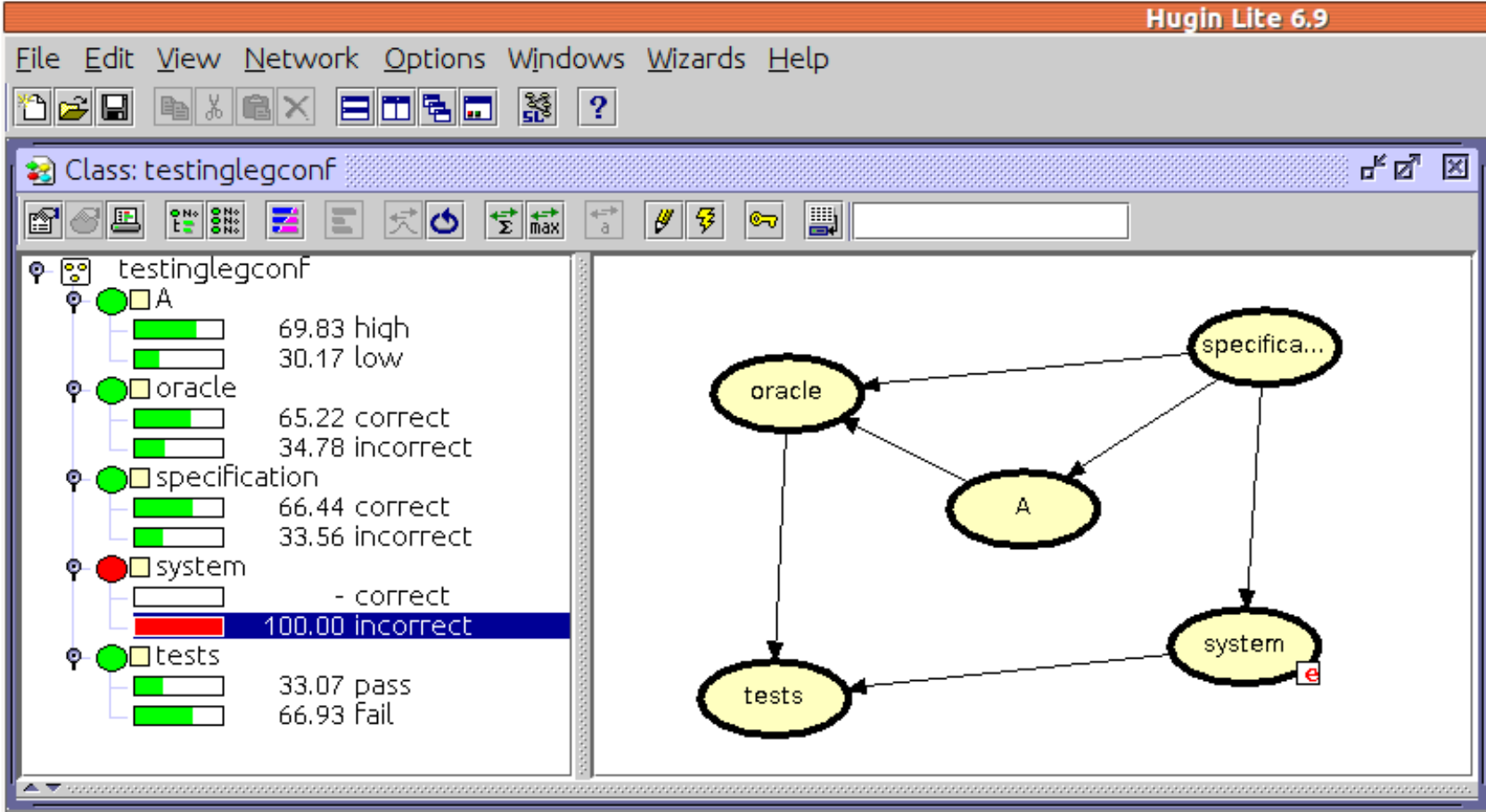}
\end{center}
\caption{\label{hugineg}Hugin Analysis of BBN for Testing Evidence}
\end{figure}

\begin{figure}[ht]
\vspace*{-4ex}
\begin{center}
\includegraphics[width=1.0\textwidth]{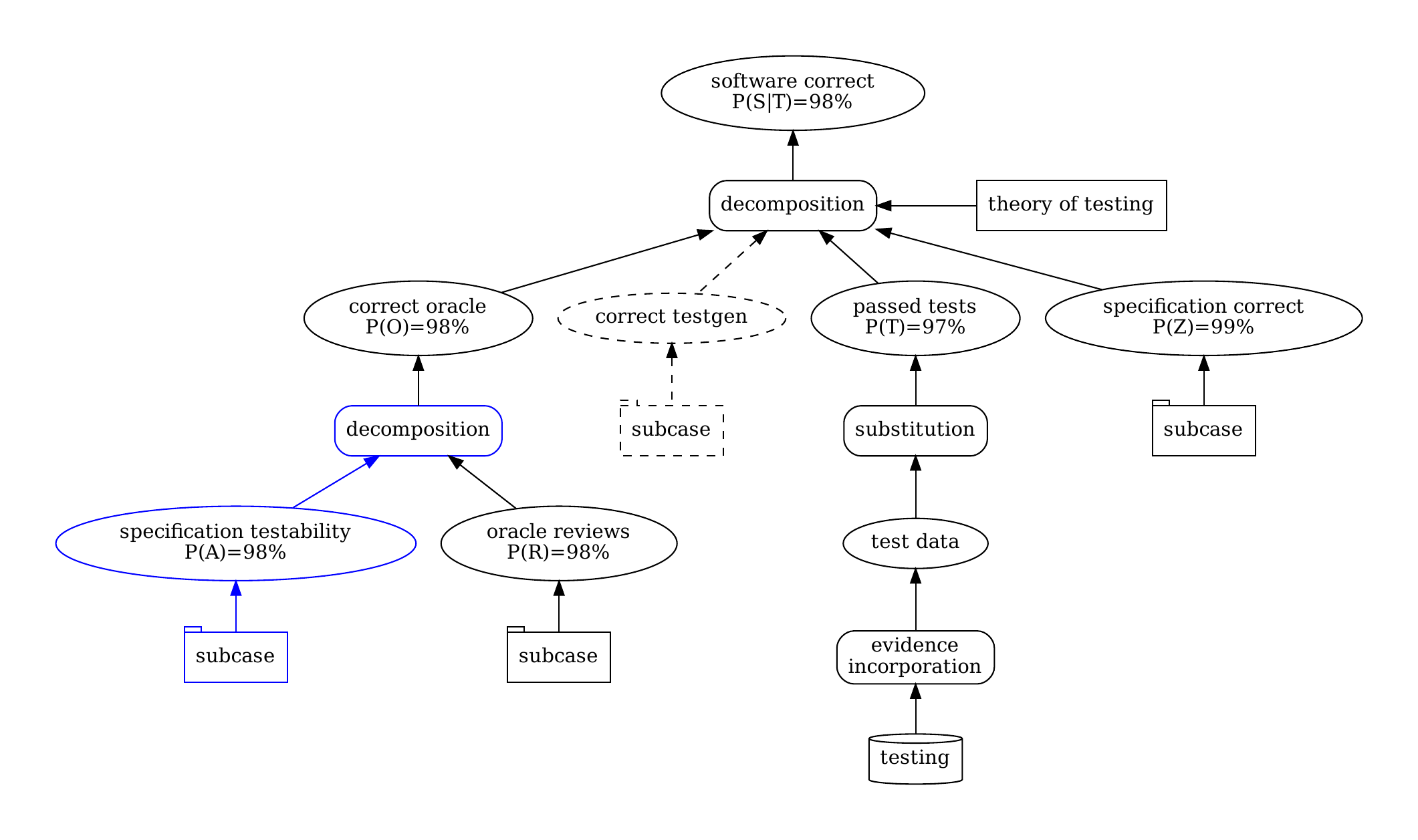}
\end{center}
\vspace*{-6ex}
\caption{\label{bbn-v2}Revised Argument for Correctness by Testing}
\vspace*{-1ex}
\end{figure}

A version of the BBN without test generation is shown represented
inside the BBN tool \emph{Hugin Expert} \cite{hugin} in Figure
\ref{hugineg}.  Here, the node labeled \texttt{specification}
corresponds to $Z$, the random variable representing correctness of
the system specification.  It has two possible values:
\texttt{correct} (i.e., it correctly represents the informal
requirements established for the system) or \texttt{incorrect}.  The
assessor must attach some prior probability distribution to these; we
will suppose 99\% confidence that it is \texttt{correct}, vs.\ 1\%
that it is \texttt{incorrect}.

The node labeled \texttt{system} corresponds to $S$, the variable that
represents the true (but unknown) quality of the system, stated as a
probability of failure on demand (that is, failure wrt.\ the informal
requirements).  This probability depends on $Z$ and is recorded in a
joint probability table (we give an example below): we suppose that it
is 99\% if $Z$ is \texttt{correct}, but only 50\% if it is
\texttt{incorrect}.

The node labeled \texttt{oracle} corresponds to $O$, the variable that
represents correctness of the test oracle; this is derived in some way
from the specification $Z$ and its probability distribution will be
some function of the correctness of $Z$; if $Z$ is \texttt{correct},
we will suppose it is 95\% probable that $O$ is \texttt{correct}, but
if $Z$ is \texttt{incorrect}, then it is only 2\% probable that $O$ is
\texttt{correct}

Finally, the node labeled \texttt{tests} corresponds to $T$, the
Boolean variable that represents the outcome of testing.  It depends
on both the oracle $O$ and the true quality of the system $S$.  Its
probability distribution over these is represented by a joint
probability table such as the following, which gives the probabilities
that the test outcome is judged successful.

\begin{center}
\begin{tabular}{|c|c|c|c|c|}
\hline
S & \multicolumn{2}{|c|}{Correct System} & \multicolumn{2}{c|}{Incorrect System}\\
\hline
O & Correct Oracle & Bad Oracle & Correct Oracle & Bad Oracle\\
\hline
T & 100\% & 50\% & 5\% & 30\%\\
\hline
\end{tabular}
\end{center}

Using a BBN tool such as Hugin, it is possible to conduct ``what if''
exercises on this model.  In particular, Hugin allows the user to
manipulate the values of some variables and observe the impact on
others.  In Figure \ref{hugineg}, we have hypothesized the system is
incorrect (indicated by the red bar, and set by double-clicking on the
value) and can see that the conditional probability that testing
succeeds (which was earlier denoted abstractly as $P(E \vbar
\neg\, C)$) is 33.07\%.

If the system is assumed correct, the probability that testing
succeeds (i.e., $P(E \vbar C)$) is 98.53\%.  We can also examine the
probability of a correct system, given that testing succeeds (i.e.,
$P(C \vbar E)$, the usual measure of assurance), which evaluates to
99.49\%, or given that it fails (i.e., $P(C \vbar \neg\, E)$), which
is 59.21\%.

We see that in this model the assumed prior distributions are such
that testing has rather poor evidential weight: it is disappointingly
likely that an incorrect system will be accepted and that a rejected
system is in fact correct.  Further inspection and experimentation
will show that part of the explanation is that the modeled test oracle
is of low quality.  The variable $O$ has strong impact on the test
outcome $T$ but is not itself observed or evaluated (in Figure
\ref{bbn-eg} it is simply assumed).  We might suppose that reliability
of the testing procedure would be improved if we could assess the
quality of the test oracle and require this to exceed some threshold.
However, it is not easy to see how this artifact can be assessed
directly, so an alternative might be to assess the ``testability'' of
the specification $Z$ (meaning the likelihood that tests derived from
Z will reveal its correctness wrt.\ the informal requirements), since
this surely has a large impact on the quality of the oracle.
Reasoning similar to this may implicitly underlie some of the DO-178C
guidelines for software assurance in civil aircraft \cite{DO178C}: for
the most critical software, DO-178C specifies 71 assurance
``objectives'' that must be accomplished and several of these concern
attributes of requirements and specifications.  For example, its
Section 6.3.2.d specifies the objective to ``ensure that each
low-level requirement can be verified.''

We can introduce this idea into our model as the variable $A$ in
Figure \ref{bbneg} (with dependencies indicated in blue) and similarly
in Figure \ref{hugineg}.  Here $A$ assesses confidence that the
specification $Z$ is testable and takes values \texttt{high} and
\texttt{low}; we suppose the probability that $A$ is \texttt{high} is
95\% when $Z$ is \texttt{correct} and 20\% when it is
\texttt{incorrect}.  There is no arc from $A$ to $S$ because $A$ is
not a general evaluation of the specification, just its testability.
The probability distribution of $O$ will now depend on both $A$ and
$Z$ and we suppose it takes the following form.

\begin{center}
\begin{tabular}{|c|c|c|c|c|}
\hline
Z & \multicolumn{2}{|c|}{Correct Specification} & \multicolumn{2}{c|}{Incorrect Specification}\\
\hline
A & High Conf & Low Conf & High Conf & Low Conf\\
\hline
O & 99\% & 70\% & 2\% & 1\%\\
\hline
\end{tabular}
\end{center}

If we require that $A$ is \texttt{high} before we undertake testing,
then we find that the probability of accepting an incorrect system is
reduced from 33.07\% to 13.33\% while the probability of accepting a
correct system increases from 98.53\% to 99.45\%.  The probability the
system is correct, given that testing succeeds, improves from 99.49\%
to 99.85\% and, if testing fails, the probability the system is
correct reduces from 59.21\% to 36.33\%.

Of course, we could informally apply similar reasoning to the argument
of Figure \ref{bbn-eg} and derive the revised argument shown in Figure
\ref{bbn-v2}.  As in the BBN analysis, we can observe that confidence
in the test oracle is dragging the overall confidence down and we can
add an evaluation of specification quality as shown (in blue) in
Figure \ref{bbn-v2} where correctness of the oracle, which was
previously an assumption, is now supported by a subargument (shown in
blue) over testability of the specification and reviews of the oracle
(recorded as the variable R).  The node labels of Figure \ref{bbn-v2}
include confidence measures, where those for the two decomposition
blocks are evaluated using the averaging method (and those for their
subclaims are simply assumed).

The primary difference between confidence propagation in the assurance
argument and the BBN analysis is that the former treats the subclaims
as if they are independent, while the latter explicitly models their
dependencies.  Both can reveal sources of weakness in an argument but
a BBN allows this to be modeled more accurately and allows calculation
of more probabilistic quantities than just propagation.

In this example, the probabilities and distributions used were
``plucked from the air'' and cannot be considered realistic.  But in a
real case, estimating probability values and tables forces serious
consideration of the interdependencies among subclaims.  However, it
is not necessarily the actual values used and calculated that are
important but the relationships among them.  We believe that
``what-if'' explorations such as those performed here can develop
understanding of these relationships and thereby help guide selection
of subcases and evidence.

Because BBNs model relationships that are absent from assurance
arguments, there is no automatic way to derive a BBN from an argument:
it requires a creative act (compare Figures \ref{bbn-eg}, \ref{bbneg},
and \ref{bbnsimp} for example).  Accordingly, we recommend that
confidence propagation in assurance case arguments is used as the
default method for quantified assessment, but that BBN analysis should
substitute or augment basic propagation through those decomposition
blocks whose subclaims have complex dependencies (simply drawing the
corresponding BBN should reveal when this is so).  Note that BBNs can
also include the sideclaim, which for simplicity we have assumed here
is treated separately.

Situations where several items of argument or evidence support a
single claim are important.  One approach assumes the separate
subcases are independent and treats them as elements of a diversity
argument; another explicitly models their dependencies in a BBN,
as Littlewood and Wright do for the combination of testing and formal
verification \cite{Littlewood&Wright:tse07}.  A third approach, which
we advocate but will not otherwise discuss here, is to develop a
separately justified theory for a specific combination of evidence:
for example, MC/DC unit testing is known to detect certain fault
classes \cite{Kuhn99:TOSEM} and we could imagine a theory that
combines this with a method of static analysis that detects other
fault classes.

\section{Comparison With Other Methods}
\label{compare}

There is much work on quantified assessment of assurance cases.
Graydon and Holloway
\cite{Graydon&Holloway:quant17,Graydon&Holloway:quant16} provide a
comprehensive description and evaluation of 12 different methods, of
which six are based on Dempster-Shafer Theory (DST) or some other form
of Evidential Reasoning, five are based on BBNs, and one uses weighted
averages of ``attributes.''

Here, we sketch some of the methods examined by Graydon and Holloway,
and highlight our differences in approach and calculation.  In the
section after this we outline the shortcomings identified in these
methods by Graydon and Holloway and consider whether they apply to our
method.

Most of the other methods start from a structured assurance argument
similar to us, though expressed in different notations, and we will
consider them first (the remaining methods start from a BBN).  Their
methods of quantification are based on probabilistic modeling and seem
to agree (among themselves and with us) in the way they deal with
assumptions and with argument nodes having a single subcase.  Where
they differ is in their treatment of evidence nodes and argument
nodes with multiple subclaims.  The differences can be in basic
analysis of these nodes (e.g., which circumstances are
recognized and treated specially) and/or the probabilistic modeling
used to represent them (e.g., different applications of DST)

We note that none of the other methods have residual \doubts or
sideclaims in their arguments, nor do they see argument steps as
applications of a theory (where sideclaims assess whether the theory
is suitable and is used correctly).  And several other methods use
notations such as GSN (Goal Structuring Notation)
\cite{GSN:community-std3} that look superficially similar to Assurance
2.0 but have different semantics.  Furthermore, the other methods do
not separate logical and numerical/probabilistic assessments: their
numerical assessments have to determine basic soundness of the case as
well as purposes specifically served by quantification.

We begin our comparison with an outline of DST and some methods that
use it.

\subsection[Methods Based on Evidential Reasoning]{Methods Based on Dempster-Shafer Theory
\newline and Other Forms of Evidential Reasoning}

Dempster-Shafer Theory was primarily developed as a theory of
evidence where we have epistemic uncertainty on the interpretation
of our data.  For example, we may wish to determine whether a
signal light is red, yellow, or green, but the best we can do with
our noisy sensors is assess a belief \emph{mass} (subjective
probability) of so much for red, so much for yellow, and so much
for green, and also so much for the weaker judgments that it is
\emph{either} red or yellow, red or green, and yellow or green,
and also so much that it could be \emph{any} of the three colors.
In other words, we assign belief masses to the powerset of the set
of basic judgments.  The \emph{belief} in a judgment is then the
sum of the mass of its subsets, so the belief for "red or yellow"
is the mass for that combination, plus those for red and yellow
individually.  In addition to belief in a judgment, we have its
\emph{plausibility}, which is 1 minus belief in its negation, so
that the plausibility of red is 1 minus belief in ``yellow or
green,''

If we have evidence from several sources, then we will wish to
combine their masses and associated functions into a consensus.
Dempster's \emph{rule of combination} assumes the separate sources
are independent and the combined belief in a judgment (e.g., light
is green) is the sum of the individual (positive) beliefs for that
judgment from the different sources multiplied by a
\emph{correction factor} related to conflicting beliefs (i.e., its
plausibility for each source).  Many different combination rules
have been proposed since the original DST
\cite{Sentz&Ferson:DScombination02} and a variety of these are
used by the methods that Graydon and Holloway consider.

Ayoub and colleagues \cite{Ayoub-etal:confidence13} divide the
different argument steps in a similar way to us.  For evidence they
use a ``basic probability assignment'' similar to our $P(C \vbar E)$,
but do not have the separation into measured and useful claims, nor do
they have sideclaims.  Nor do they use anything comparable to
confirmation measures to assess the epistemic significance and
relevance of evidence.

Their approach to reasoning nodes that have a single subclaim is
similar to ours, except they do not apply a function \texttt{f}, nor
do they have sideclaims.  Where Ayoub and colleagues differ most from
us is in probabilistic modeling of decomposition steps, where they use
DST\@.  They divide cases in a similar way to us: what we call
diversity, they call an ``alternative'' argument, and what we call
partitioning, they call a ``disjoint'' argument.  We will use our
terminology.  For partitioning arguments, they favor a ``weighted
mixing'' combination rule which evaluates similarly to our
partitioning calculation, but for diversity arguments they revert to a
standard DST combination rule and obtain very different results to us.
They also have containment arguments and other cases that they reduce
to diversity and partitioned arguments in ways similar to our Sections
\ref{containment} and \ref{others}.  However, their numerical results
usually differ from ours because their treatment of the underlying
diversity cases is different to ours,

The method of Zeng, Lu, and Zhong \cite{Zeng&Lu&Zhong13} is similar to
that just considered except that the subclaims to all nodes are
assigned a weight as well as a confidence (like our partitioned cases)
and are combined using an ``improved Dempster's Rule.''

Cyra and G\'{o}rski \cite{Cyra&Gorski11} present a method for
``Trust-IT'' arguments that are rather different to the arguments of
Assurance 2.0, so detailed comparison is difficult.  In particular,
they can have counterevidence or counterarguments present in the final
case, whereas we regard these as defeaters that may be used to
challenge the case during development and evaluation, but must be
resolved in the final case, which is entirely positive.

They use J{\o}sang's Opinion Triangle \cite{Josang16} to evaluate
evidence and other leaf nodes.  The Opinion Triangle considers belief,
disbelief, and uncertainty and has similar motivation, but different
calculations, to the confirmation measures that we use.  The resulting
assessments are converted into DST belief and plausibility measures
and these are propagated up the argument, step by step, using DST with
a qualitative probability scale that combines ``decision'' and
``confidence'' to yield a 24-point scale.

They divide argument steps into four kinds and use different DST
combination rules for each.  Their ``complementary'' arguments are
like our partitioned cases but may have a ``gap'' that is not covered
by any subcase (which we would absorb into the sideclaim).
Quantification for this case uses a modified DST combination rule
rather similar to that of Ayoub and colleagues.  Their ``alternative''
arguments correspond to our diversity cases and are quantified using
Yager's \cite{Sentz&Ferson:DScombination02} modified DST combination
rule.  Their ``Necessary and Sufficient Condition List'' arguments
resemble our logical assessment of decomposition blocks, while their
``Sufficient Condition List'' arguments are similar except that not
all the subclaims are necessary (so we might combine the
``overlapping'' subclaims into a diversity subcase).  Both of these
are quantified using novel DST combination rules.

Like Cyra and G\'{o}rski, Duan and colleagues
\cite{Duan-etal:confidence15} use J{\o}sang's Opinion Triangle, but
they apply it to the entire argument, not just the evidence nodes.  In
other words, they use J{\o}sang's method for evidential reasoning in
place of DST\@.  The method represents degrees of belief, disbelief,
and uncertainty according to beta distributions and has formulas for
combining multiple assessments similar to the rules of combination in
DST\@.  Like DST, we consider this method ill-suited to quantifying
confidence in Assurance 2.0, where our goal is to assess the strength
of a logically sound argument, not to assess whether it is sound.

Guiochet, Hoang, and Ka\^{a}niche \cite{Guiochet-etal14:confidence}
introduce a method based on DST that uses modified rules of
combination so that its treatment for diversity and partitioned cases
is arithmetically similar to ours except that it has some additional
parameters.  However, when an argument node has a context node
attached (they assume GSN arguments), the analysis instead uses a BBN
model similar to Figure \ref{bbnsimp}.  They treat context nodes
rather like our sideclaims, and this is criticized by Graydon and
Holloway who remark that in GSN ``context is not a proposition''
\cite[page 59]{Graydon&Holloway:quant16}.

Nair and colleagues \cite{Nair-etal:ER15} use a method for evidential
reasoning due to Yang and Xu \cite{Yang&Xu02} that is different to
DST, but related to it.  They do not apply evidential reasoning to the
assurance case itself, but to a separate assessment of confidence in
the case (a ``confidence argument'' \cite{Hawkins-etal:SSS11}).  Their
constructions are somewhat specific to GSN and at variance with
Assurance 2.0.  In particular, probabilistic assessment in Assurance
2.0 is applied only to assurance arguments that are judged to be
indefeasibly sound, whereas these assessments are combined in the
method of Nair and colleagues.

At the beginning of this section we stated that evidential reasoning,
whether based on DST, J{\o}sang's Triangle, or other methods, is an
uncomfortable fit for Assurance 2.0 and we believe this is confirmed
by the methods sketched above.  As the term ``evidential reasoning''
suggests, these methods are plausible for the evaluation of evidence
with respect to claims (where we use confirmation theory), although
Dempster's rule of combination has been called into question
\cite{Tchamova&Dezert12,Dezert-etal:DST12} and the plethora of
alternative rules indicates the details of its application have not
been settled satisfactorily.  But once we have incorporated evidence
into claims, the interior part of an assurance case argument is about
logical reasoning and this is poorly, not to say inaccurately,
represented by DST and similar methods.  We believe this is because
these methods are attempting to perform three different assessments
simultaneously: logical soundness, dialectical examination, and
evaluation of argument strength.  In assurance 2.0, we perform these
assessments separately.  Dialectical examination is performed with
defeaters, counterarguments and counterevidence.  Only once these have
been resolved do we proceed to logical soundness, where we have only
positive arguments and can employ standard logical reasoning.  And
only once soundness is confirmed do we assess the strength of the
argument, using elementary probabilistic modeling.

In conclusion, we do not consider DST and other forms of evidential
reasoning to be applicable or to add value to Assurance 2.0 and so we
next turn to methods based on BBNs.

\subsection{Methods Based on Bayesian Belief Networks}
\label{altbbns}

In Section \ref{bbns} we described the use of BBNs to examine the
consequences of dependencies among the subclaims of a decomposition
block within an assurance argument.  The BBN was constructed by human
evaluators and served to augment and refine the standard confidence
assessments for an assurance argument.

Among the BBN-based methods considered by Graydon and Holloway, those
of Guo \cite{Guo:BBNs03}, Hobbs and Lloyd \cite{Hobbs&Lloyd11}, and
the SERENE Partners \cite{SERENE-method99} use BBNs of a similar form
to those in Section \ref{bbns} but construct them directly, without
first developing an explicit assurance argument: in effect, the BBNs
\emph{are} their representation of the argument.  Because they do not
relate their methods to assurance arguments as we understand them in
Assurance 2.0, we find their methods remote from our concerns and do
not consider them further.

\begin{figure}[t]
\begin{minipage}[t]{2.9in}
\vspace*{0.05in}
\includegraphics[width=2.5in]{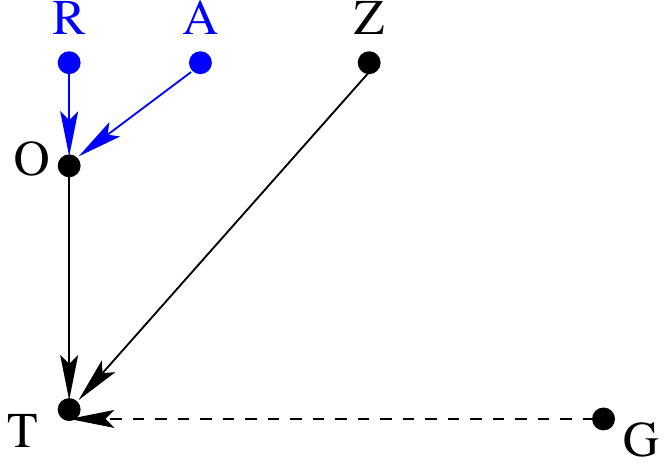}
\end{minipage}\hfill\begin{minipage}[t]{2.3in}\raggedright
\vspace*{0.03in}
\begin{description}\itemsep=1ex
\item[\textbf{Z}:] System Specification
\blue{\item[\textbf{A}:] Specification quality}
\blue{\item[\textbf{R}:] Oracle reviews}
\item[\textbf{O}:] Test Oracle
\item[\textbf{T}:] Test results
\item[\textbf{G}:] Test generator
\end{description}
\end{minipage}
\caption{\label{bbnsimp}BBN Derived from Assurance Argument for Testing Evidence}
\end{figure}

The method of Denney, Pai, and Habli \cite{Denney-etal:confidence11}
does base its BBN on an assurance argument, but does so in an
automated fashion, As we explained in Section \ref{bbns}, the
dependencies among subclaims that are modeled and analyzed by BBNs are
not represented in assurance arguments and so a BBN of this form
requires human insight and cannot be derived automatically from the
argument.  Consequently, the BBNs of Denney, Pai, and Habli have an
impoverished form as illustrated in Figure \ref{bbnsimp}, which is
derived from Figure \ref{bbn-v2}.  This BBN has an arc from each
(variable associated with a) claim to its parent, and is therefore
isomorphic to the argument structure (with its arrows reversed, and
drawn upside-down).  Denney, Pai, and Habli do add a random variable
``assurance deficits acceptable,'' supported by variables ``argument
sufficient'' and ``context appropriate'' to the BBN, but these are
unrelated to the explicit argument structure.  The variables of the
BBN are discretized probabilities with normal distributions assumed
(presumably to simplify the calculations).  It seems to us that this
method introduces the complexity of BBNs while lacking their chief
merit: the ability to model dependencies among subclaims to
decomposition nodes.

Zhao and colleagues \cite{Zhao-etal-confidence12} convert an assurance
case argument into a sequence of Toulmin arguments
\cite{Toulmin:uses-of-argument} and then build a BBN for each.  Their
BBNs employ standard patterns and thereby fail to model dependencies
among subclaims.  We consider the excursion into Toulmin arguments and
their method of BBN construction to be sufficiently different in
philosophy and technique to ours that we do not find their techniques
applicable to our concerns.

\subsection{Other Methods}

Yamamoto \cite{Yamamoto-confidence15} describes a method for
evaluating ``attributes'' of a GSN assurance argument.  Attributes are
not probabilities nor estimates of confidence; they are judgments
about the truth of claims, represented as integers on a five-point
scale, where -2 corresponds to ``strongly unsatisfied,'' -1 to
``unsatisfied,'' 0 to ``unknown,''  1 to ``satisfied'' and 2 to
``strongly satisfied.''

He uses a calculation for GSN strategies that resembles our
partitioned case, but provides no justification.  His description is
terse and difficult to understand.  We reproduce it here in full
\cite[Section IV\!, A]{Yamamoto-confidence15}:
\begin{quote}
2) Sub claim to parent claim propagation through strategy

``Let a parent claim is decomposed by $k$ sub claims through a
strategy.  And let $<\!\!<\!P\!>\!\!>$ be the attributes of the
parent. Let $<\!\!<\!Q_1,\ldots,Q_k\!>\!\!>$ be the attribute of the
strategy. Let $<\!\!<\!R_1\!>\!\!>,\ldots,<\!\!<\!R_k\!>\!\!>$ be the
attributes of $k$ sub claims, respectively.

``Then the attribute value $P$ of the parent claim is calculated
by the following equation.

$P = (\Sigma_{i=1,k}\ Q_i \times R_i \times W_i)$, where
$\Sigma_{i=1,k}\ W_i = 1$''
\end{quote}
Earlier, he explains that the $W_i$ are weights:
\begin{quote}
``$W_{1}, \ldots, W_k$, where $k$ is the number of sub claims of the
corresponding strategy. $W_i$ is the weight for the $i$'th sub
claim.''
\end{quote}

It is not clear how to interpret the $Q_i$ and they are not used in
the case study \cite[Section V]{Yamamoto-confidence15}, so we will
ignore them and the formula is then \[P = \Sigma_{i=1,k}\ R_i \times
W_i,\] which is the same as our formula for partitioned decomposition
blocks (\ref{parteqn}) stated in Section \ref{parts}.

However, although Yamamoto's arithmetic calculation is the same as
ours, he applies it to different quantities (``attributes'' rather
than probabilistic confidence) and provides no justification.
Furthermore, he does not stipulate that it applies only to partitioned
decompositions and does not consider the diversity case for
decompositions.  Consequently, we regard our method as distinct from
his.

\section{Graydon and Holloway's Critique}
\label{GandH}

For many of the 12 methods they examine, Graydon and Holloway
reproduce the authors' examples and then construct variants that
generate implausible results.  Many of the variants are based on
similar ideas, so we next consider those.  We reference the technical
report \cite{Graydon&Holloway:quant16} as its appendices give more
details than the paper \cite{Graydon&Holloway:quant17}, and we use
paragraph headings taken from the report.

\paragraph{Many Subclaims Examples \cite[pp.\
36--39]{Graydon&Holloway:quant16}.}  Graydon and Holloway develop this
group of examples for the method of Ayoub and colleagues
\cite{Ayoub-etal:confidence13} but provide similar constructions for
several other methods.  Abstracting from the details, the idea is that
we have a decomposition step with many subcases, say 20.  Of these, 19
have high confidence and one is very low.  Most methods, including
those introduced in Section \ref{method}, are not particularly
sensitive to the degree of confidence in one subclaim among many, so
confidence in the parent claim will be dominated by the majority and
will therefore be high in this example.  If acceptability of the
argument is based on numerically high confidence, then this argument
will be accepted despite having one very weak element, and Graydon and
Holloway rightly regard this as a cause for concern.

But this is not how Assurance 2.0 works, and the example is no threat
to our methods.  In Assurance 2.0, we do not perform probabilistic
assessment until we have determined that the argument is logically
sound, which means each parent claim must be deductively entailed by
its subclaims.  The subclaim with low confidence would surely not have
been adjudged \texttt{true} in logical assessment and so its parent
claim and all those above it (including the top claim) will be
\texttt{false} or \texttt{unsupported} and this assessment must pause
while the case is examined and corrected.

On the other hand, logical assessment will not highlight the case
where subclaims are plausibly judged \texttt{true} but with less
certainty than others: it does not even provide a way to talk about
this.  Probabilistic assessment in Assurance 2.0 is a lens that gives
a different view of the case and allows us to speak rationally of
confidence, so that weakly \texttt{true} subclaims that might
otherwise have gone unremarked will be identified and can then lead to
further investigation.

Graydon and Holloway consider variants on this scenario where
a subclaim provides undermining or counter-evidence but is
overwhelmed by other, positive, subclaims so that the overall
assessment is positive.  These circumstances can arise in
Yamamoto's method and also in those using DST and other evidential
reasoning, where quantification extends to disbelief as well as
belief.  But in Assurance 2.0 we examine disbelief using defeaters
and do this---and adjust matters to rectify serious sources of
\doubt---prior to probabilistic assessment, where the case is
entirely positive and the proposed scenarios cannot arise.

The examples with many subclaims also raise the question of arbitrary
scope: why do we have these subclaims and not some other number and
selection?  Graydon and Holloway consider methods that perform
calculations similar to our partitioning case (Section \ref{parts})
and suppose we have two subclaims with confidence 95\% and 50\%,
respectively.  The averaging variant of our partitioning case will
calculate confidence in the parent claim as 72.5\%.  Now suppose we
replace the first subclaim by 10 smaller subclaims, all with the same
95\% confidence.  The averaging method now calculates the parent
confidence as 90.9\%.

Graydon and Holloway criticize this example's sensitivity to an
arbitrary choice in how to divide subclaims, but we consider it
contrived.  The idea of a partitioning case is that the argument
naturally divides into separate subclaims, so the choice is
unlikely to be arbitrary.  And where there is some flexibility of
choice, confidence in the subclaims will vary appropriately and
the parent confidence should not be distorted.  As a trivial
example, we could replace evidence from a test campaign with
evidence from the same campaign divided into 10 parts---but
confidence in each smaller campaign will be less than that in the
undivided whole, so the parent confidence will not change wildly
but appropriately.

\paragraph{Imperfect Examples \cite[page
45]{Graydon&Holloway:quant16}.}  Graydon and Holloway provide
several examples that challenge the method of Cyra and G\'{o}rski.
Essentially, one of Cyra and G\'{o}rski's formulas calculates
confidence in a parent claim as something similar to the product
of confidence in its subclaims.  Consequently one subclaim of
slightly lower confidence than the others will sharply lower
confidence in the parent claim; confidence in the parent claim is
also very sensitive to the number of subclaims.

Their method is similar to our older ``product'' calculation of
Section \ref{oldmethod} and observations such as these are the
reason we now deprecate its use except in special circumstances
(e.g., recall Section \ref{cumulative}, where it is exactly the
right method).  However, we note that calculations that are
sensitive to subclaims of lower confidence and to numbers of
subclaims, such as product and sum of doubts, can be useful
indicators of circumstances where further investigation may be
warranted, so although we no longer recommend these methods as the
primary means of probabilistic assessment, we can recommend them
as special-purpose supplements to those of Section \ref{method}.

\begin{wrapfigure}{R}{.5\textwidth}
\vspace*{-2ex}
\hspace*{-2ex}\includegraphics[width=0.55\textwidth]{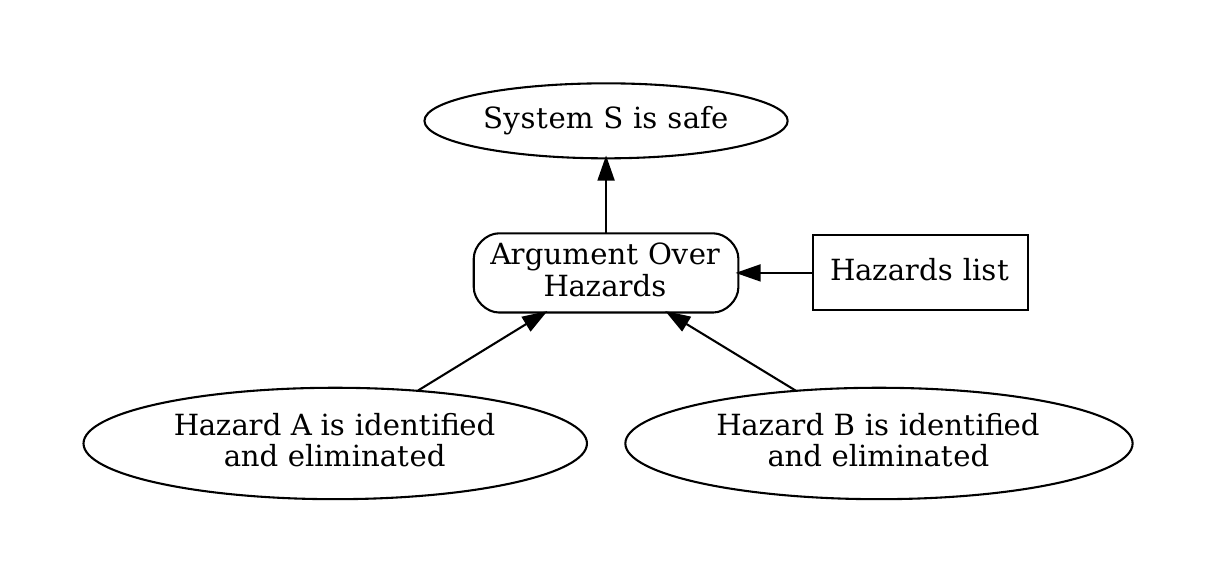}
\vspace*{-6ex}
\caption{\label{zhao-cf}Example from \cite[Fig.\ 7]{Zhao-etal-confidence12}}
\vspace*{-2ex}
\end{wrapfigure}

\paragraph{Optimistic and Pessimistic Counterexamples \cite[pp.\ 87--90]{Graydon&Holloway:quant16}.}
Graydon and Holloway consider the trivial example from Zhao
and colleagues \cite[Figure 7]{Zhao-etal-confidence12} that we
translate from GSN to Assurance 2.0 and reproduce in Figure
\ref{zhao-cf}.  They show that when confidence in the sideclaim
(that establishes completeness of the hazard list) is
changed from the original 85\% to 99.9\% and then to 0.01\%,
confidence in the top claim of system safety changes from 89.460\%
to 89.996\% and then to 86.404\%.  They rightly note that ``it is
not plausible that extreme changes in confidence in hazard
analysis would produce as small a change in confidence in
safety.''

In Assurance 2.0, the version with confidence 0.01\% would be
rejected as logically unsound, as a sideclaim with such low
confidence could not possibly be adjudged \texttt{true}.  For the
other cases, our methods of Section \ref{method} calculate
confidence in a parent claim as the product of confidence in its
combined subclaims (independently of how that is calculated) and its
sideclaim.  Zhao and colleagues assume confidence in the combined
subclaims is 90\%, so our methods deliver 76.5\% confidence in the
top claim when sideclaim confidence is 85\% and 89.9\% when it is
99.9\% (and would deliver 0.009\% if we persisted with sideclaim
confidence of 0.01\%).  We consider these to be appropriate values.

The method of Zhao and colleagues uses BBNs, which we also endorse in
suitable circumstances but, as we noted in Section \ref{bbns}, BBNs
need to be constructed to reflect the specific circumstances of the
argument under consideration.  We suggest that this example of Graydon
and Holloway demonstrates the unsuitability of generic ``one size fits
all'' BBN constructions.  Similar examples and observations apply to
other BBN-based methods and examples examined by Graydon and Holloway.

\section{Summary and Conclusion}
\label{conc}

An Assurance 2.0 argument may contain probabilistic claims,
including claims about confidence, and can employ different kinds of
probabilistic reasoning within the argument.  We distinguish four
kinds and a single assurance case may employ elements of all four.
\begin{enumerate}

\item At one extreme there are no probabilistic claims, only
unconditional claims such as ``implementation is correct wrt.\
specification'' supported by similarly unconditional reasoning.

\item Next, we may have qualitative claims about stochastic
properties such as ``the new system will be no less reliable than
the old'' with limited evidence and reasoning about relative
reliability.

\item Then, we may have quantitative claims about stochastic
properties such as ``reliability is better than $x$'' supported by
evidence, theories, and reasoning about these properties.

\item Finally, we may have claims explicitly about confidence, such
as ``95\% confidence that reliability is better than $x$'' supported
by suitable probabilistic theories and reasoning.

\end{enumerate}

The last three of these include various amounts and degrees of
numerical and probabilistic reasoning that is internal to (i.e.,
part of) the argument.  In addition, the first three can be
supported by probabilistic reasoning that is external to (i.e.,
separate from) the reasoning in the argument (so can the fourth, but
it seems redundant).  This external reasoning assesses
\emph{confidence} in each claim: that is, a subjective probabilistic
assessment of its truth taking aleatoric and epistemic uncertainties
into account.  This is different to the logical assessment of truth,
which assesses validity and soundness of the argument (given that
all evidence and all reasoning steps are justified indefeasibly).

Probabilistic confidence assessment in Assurance 2.0 is performed only
for arguments that have been judged logically sound.  \Doubts are
explored by dialectical examination using defeaters and is a separate
(prior) activity to logical and probabilistic assessment, which are
applied only to strictly positive cases in which any residual \doubts
are explicitly noted and accepted.  The value added by probabilistic
confidence assessment is to supplement dialectical and logical
assessments (by providing additional information or a different point
of view), to enable one argument to be compared with another, to
examine allocation of effort across an argument, and to support
rational tradeoffs of effort against risk for graduated levels of
assurance.

An example supplementation is ``chain of confidence'' reasoning
where we model the impact of being wrong, and how wrong we might be
\cite[Annex I 1.5]{IAEA18:nuclear}.  For example, we may estimate
$R_\ok$ as a measure of risk posed by the system, given the
assurance argument.  But if $P_\ok$ is the probability our assurance
argument is correct (i.e., our confidence in it), and $R_\neg\ok$ is
an estimate of risk absent this argument, then the chain rule gives
\[R = P_\ok \times R_\ok + (1-P_\ok) \times R_{\neg\ok}\]
and we can use this to estimate a refined measure of overall risk $R$.

In this report, we have described a new method for calculating
confidence in the claims of an Assurance 2.0 argument.  The method
proceeds one argument step after another, propagating confidence from
leaf nodes such as evidence, assumptions, and residual \doubts, up to
the top claim.  At the leaves, confidence in evidence \texttt{E} is
estimated as \texttt{P(C\vbar E)} where \texttt{C} is its ``something
useful'' claim.  Confidence in assumptions is estimated by subject
matter experts or other assessors.  Confidence for residual \doubts is
an estimation of confidence that they are truly residual (i.e., pose
only negligible risk).

At each argument step above the leaves, the method estimates the
quantity of \doubts in the parent claim that are eliminated by its
subclaims.  For decomposition steps, the assessment depends on the
reasoning concerned: for example, do the subclaims partition the
\doubts, or do they eliminate \doubts in diverse ways?  Despite their
differences, these assessments are all derived from the same
\doubt-elimination model and are easy to understand, to justify, and
to calculate.  In some circumstances, where subclaims have complex
interdependencies, we endorse use of BBNs, but these must be
carefully crafted to represent the dependencies concerned and cannot
use the generic forms proposed for some other methods.  For most
purposes, however, the complexity of an ad-hoc BBN is unnecessary: a
specialized theory for the relevant dependencies (e.g., for testing
used in combination with static analysis) will be preferable, while
a conservative approximation using one of our basic constructions
will often suffice.

At each argument step, calculated confidence in the parent claim is
multiplied by confidence in its sideclaim, if there is one (because
truth of the parent follows from the subclaims \emph{given} the
sideclaim), and ad-hoc adjustments, represented by a function
\texttt{f} or a product \texttt{k}, may be applied if warranted.

Graydon and Holloway examine several methods for probabilistic
assessment of assurance cases and show that all of them can generate
implausible results
\cite{Graydon&Holloway:quant17,Graydon&Holloway:quant16}.  Most of
these arise because the methods do not assess logical soundness
separately from their probabilistic calculations, which can therefore
be applied to unsound arguments.  Our method is not susceptible to
these or other counterexamples developed by Graydon and Holloway.

Several experiments, surveys, and interviews have attempted to explore
users opinions and experiences in evaluating assurance cases.  Graydon
and Holloway mention (and criticize) a couple and Diemert, Shortt, and
Weber describe several more and present their own
\cite{Diemert-etal:conf25}.  Few of the participants in these reviews
show much enthusiasm for quantitative methods.  Diemert and colleagues
report skepticism that nuanced assessments can be ``reduced to a
number'' and also doubts that quantitative methods produce trustworthy
results.  We hope that the role we propose for quantitative methods in
Assurance 2.0, where they complement reviews and logical and
dialectical assessments, and the intuitive and simple calculations
that underlie our methods, can allay these doubts and concerns and
will prove valuable when using Assurance 2.0.

One of the roles where we propose that our quantitative methods may be
particularly useful is in evaluating tradeoffs in effort \emph{vs}.\
confidence when developing graduated assurance.  Daw, Beecher, and
Holloway discuss these tradeoff issues \cite{Daw-etal:levels-DASC23}
and it would be interesting to compare our quantitative methods with
their informal approaches.

\paragraph{Acknowledgment and Disclaimer.}\mbox{}\\

We thank Peter Bishop, Bev Littlewood, and Andrey Povyakalo of City St
George's for their challenging critique of some of our description
and derivations.

This material is based upon work supported by the United States Air
Force and DARPA under Contract No. FA8750-23-C-0519 and is released
under Distribution Statement ``A'' (Approved for Public Release,
Distribution Unlimited).  Any opinions, findings and conclusions or
recommendations expressed in this material are those of the author(s)
and do not necessarily reflect the views of the United States Air
Force, DARPA, or the United States Government.

\section*{References\markboth{References}{References}}
\addcontentsline{toc}{section}{References}
\bibliographystyle{modplain}

\end{document}
% End of foo.tex -----------------------------------